\documentclass[oneside]{ar-1col_hacked}
\usepackage{url}
\usepackage[numbers]{natbib}
\usepackage{amsmath,amssymb}
\usepackage{hyperref}
\usepackage{booktabs}
\usepackage{orcidlink}
\usepackage[left]{lineno}

\jname{Xxxx. Xxx. Xxx. Xxx.}
\jvol{AA}
\jyear{YYYY}
\doi{10.1146/((please add article doi))}

\begin{document}

\markboth{Akitaka Ariga et al.}{Neutrino Experiments at the Large Hadron Collider}

\title{Neutrino Experiments at the Large Hadron Collider}

\author{Akitaka Ariga \orcidlink{0000-0002-6832-2466},$^{1,2}$ Jamie Boyd \orcidlink{0000-0001-7360-0726},$^3$ Felix Kling \orcidlink{0000-0002-3100-6144},$^4$ and Albert De Roeck \orcidlink{0000-0002-9228-5271} $^3$ \\ \\
\small{\textmd{$^1$ {Albert Einstein Center for Fundamental Physics, Laboratory for High Energy Physics, University of Bern,
Sidlerstrasse 5, CH-3012 Bern, Switzerland, \texttt{akitaka.ariga@unibe.ch }}}} \\
\small{\textmd{$^2$ {Department of Physics, Chiba University, 1-33 Yayoi-cho Inage-ku, 263-8522 Chiba, Japan}}}\\
\small{\textmd{$^3$ {CERN, CH-1211 Geneva 23, Switzerland, \texttt{jamie.boyd@cern.ch}, \texttt{albert.de.roeck@cern.ch}}}}\\
\small{\textmd{$^4$ {Deutsches Elektronen-Synchrotron DESY, Notkestr.~85, 22607 Hamburg, Germany, \texttt{felix.kling@desy.de}}}}\\}

\begin{abstract}
The proton-proton collisions at the Large Hadron Collider (LHC) produce an intense, high-energy beam of neutrinos of all flavors, collimated in the forward direction. Recently two dedicated neutrino experiments, FASER and SND@LHC, have started operating to take advantage of the TeV energy LHC neutrino beam, with first results released in 2023 and further results released in 2024. The first detection of neutrinos produced at a particle collider opens up a new avenue of research, allowing to study the highest energy neutrinos produced in a controlled laboratory environment, with an associated broad and rich physics program. Neutrino measurements at the LHC will provide important contributions to QCD, neutrino and BSM physics, with impactful implications for astro-particle physics. This review article summarizes the physics motivation, status and plans of, present and future neutrino experiments at the LHC.
\end{abstract}

\begin{keywords}
neutrinos, LHC, FASER, SND@LHC, FPF
\end{keywords}
\maketitle

\tableofcontents

\section{INTRODUCTION}

The LHC is the most energetic particle collider built thus far, colliding protons with a center-of-mass energy of about 14~TeV. Its primary focus is to study rare but spectacular events in the central region, such as those expected to be caused by the decay of Higgs bosons or proposed heavy particles at the TeV scale, and the main LHC detectors were designed for this purpose. In addition, the LHC also produces the most energetic neutrinos created by humankind: proton-proton collisions generate an intense and strongly-collimated beam of highly-energetic neutrinos and antineutrinos of all three flavors in the forward direction along the collision axis. Since the neutrino interaction cross section rises roughly linearly with energy, already modestly sized detectors placed into this beam are able to collect significant event rates and therefore allow to utilize particle colliders as a neutrino source. 

The idea of detecting LHC neutrinos was first pointed out in 1984~\cite{DeRujula:1984ns} and reconsidered several times in the following decades at a conceptual level~\cite{Winter:1990ry, DeRujula:1992sn, Vannucci:1993ud, Park:2011gh, Foldenauer:2021gkm}. However, it would take until the late 2010s before the first steps towards an experimental program were made independently by two groups. In 2018, FASER (ForwArd SEarch expeRiment) was proposed to search for light feebly-interacting particles and a location in the LHC tunnel system was identified that allowed to access the collision axis~\cite{Feng:2017uoz, FASER:2018ceo, FASER:2018bac, FASER:2018eoc, FASER:2019aik}. During that time, a small prototype emulsion detector was installed to assess the background levels, and the collected data were used to report the first neutrino interaction candidates~\cite{FASER:2021mtu}. A year later, the dedicated neutrino detector FASER$\nu$ was added to the experiment~\cite{FASER:2019dxq, FASER:2020gpr}.  At the same time, a separate group of researchers investigated the feasibility of placing neutrino detectors at various locations inside the LHC tunnel system~\cite{Buontempo:2018gta, Beni:2019gxv, Beni:2020yfy}. These efforts led to the proposal of the XSEN (Cross Section of Energetic Neutrinos) detector~\cite{XSEN:2019bel}, and in a common effort with the proponents from the earlier proposed neutrino detector of the SHiP experiment, called SND~\cite{SHiP:2020sos, Ahdida:2750060}, united into the SND@LHC (Scattering and Neutrino Detector at the LHC) detector project. 

Both the FASER and SND@LHC experiments were installed during the previous long shutdown of the LHC (in the period 2019 to mid-2022) and have been successfully taking data since the beginning of LHC Run~3 in the summer of 2022. As their first achievements, both experiments reported the observation of collider neutrinos in March 2023~\cite{FASER:2023zcr, SNDLHC:2023pun}. The experiments will continue to operate until the end of LHC Run~3 in 2026. Continued operations and upgrades of both experiments are planned for the LHC Run~4 starting in 2030~\cite{Boyd:2882503, Abbaneo:2895224, Abbaneo:2909524}. In addition, an expansion of the collider neutrino physics program with significantly larger detectors during the high-luminosity LHC (HL-LHC) era has been proposed in the form of the Forward Physics Facility (FPF)~\cite{MammenAbraham:2020hex, Anchordoqui:2021ghd, Feng:2022inv, Adhikary:2024nlv} and more recently also at the surface exit points of the neutrino beam~\cite{ Ariga:2025jgv,Kamp:2025phs}. The existing and proposed collider neutrino experiments will carry out a broad physics program, to analyze neutrino interactions at the highest energy from an artificial source, investigate properties of tau neutrinos, study the strong interaction in unexplored kinematic regimes, provide important input for astroparticle physics, and search for signs of beyond the Standard Model (BSM) physics. 

\section{PHYSICS MOTIVATION}
\label{sec:physicsmotivation}

\begin{figure}[hb]
\centering
\includegraphics[width=1.\textwidth]{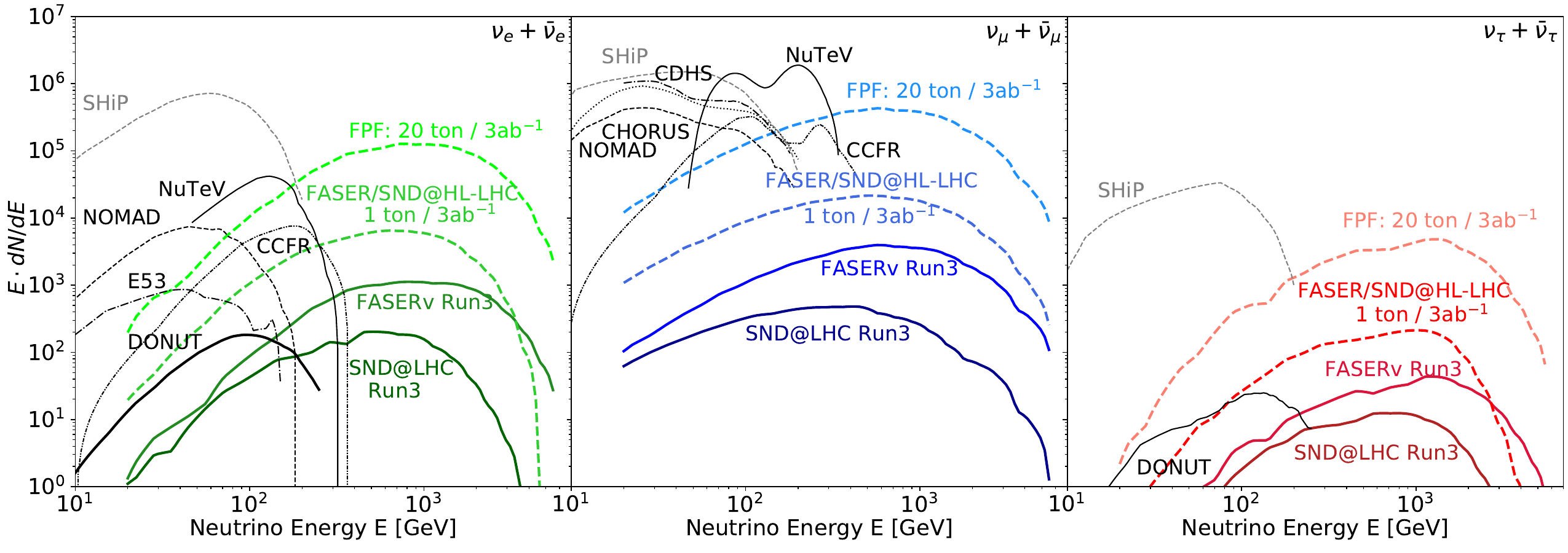}
\caption{\textbf{Neutrino yields at existing and proposed collider neutrino experiments.} The panels show the expected event rate and energy spectrum of neutrinos interacting at existing and proposed collider neutrino experiments for electron (left), muon (middle), and tau (right) neutrinos. Details of the simulation are discussed in Sec.~\ref{sec:expComp}. Spectra from previous accelerator experiments and the planned SHiP experiment are shown for comparison. Figure adapted from Ref.~\cite{Adhikary:2024nlv} (CC BY 4.0).}
\label{fig:spectra}
\end{figure}

The first observation of neutrinos at the LHC marks the \textit{dawn of collider neutrino physics}~\cite{Worcester:2023njy}. While only a few hundred neutrinos have been observed so far, the existing experiments are projected to record around ten thousand neutrinos by the end of LHC Run~3 in mid-2026. Looking further into the future, about a hundred thousand neutrinos would interact in similarly sized detectors during the HL-LHC era, while larger scale detectors proposed for the FPF could observe more than a million collider neutrinos. This is illustrated in Figure~\ref{fig:spectra}, which shows the expected event rates and energy spectra of existing and proposed collider neutrino experiments in comparison to previous accelerator experiments as well as the planned neutrino detector of the SHiP experiment~\cite{Aberle:2839677, Ahdida:2023okr}. Notably, the energies provided by the LHC neutrino beam exceed those of previous facilities. Together with the increasingly large event statistics, this offers a variety of unique opportunities to constrain physics associated with neutrino production and interactions as well as to search for signs of new physics. \smallskip

\noindent \textbf{Neutrino Interaction Measurements:} The energy of the LHC neutrino beam exceeds that of all other artificial neutrino sources and is only surpassed by atmospheric and astrophysical neutrinos. This allows the LHC neutrino experiments to perform the measurement of neutrino interaction cross sections at otherwise inaccessible TeV energies, thereby validating the neutrino cross section models and providing important input for neutrino telescopes~\cite{Candido:2023utz, Jeong:2023hwe}. This also includes exclusive channels like the charm associated neutrino interactions $\nu q \to \ell c$ and possibly even the first observation of beauty associated interactions $\nu q \to \ell b$. These measurements can be performed for all three neutrino flavors, including tau neutrinos. While only about ten tau neutrino interactions have been previously identified on an event-by-event basis by each of the DONuT and OPERA experiments~\cite{DONuT:2007bsg, OPERA:2018nar}, significantly larger numbers of events could be collected with collider neutrino experiments. This provides a novel opportunity to perform precision measurements of tau neutrino properties~\cite{MammenAbraham:2022xoc} and to probe lepton universality in neutrino scattering by comparing the interaction rates of all three neutrino flavors~\cite{Falkowski:2021bkq}. If equipped with charge identification capabilities, the collider neutrino experiments can also separately observe the tau neutrinos and tau antineutrinos for the first time through their decays into muons.

The forward neutrino experiments also probe charged-current (CC) deep-inelastic scattering (DIS) with center-of-mass energies in the range of 10 to 50 GeV, described in Ref.~\cite{Cruz-Martinez:2023sdv} as augmenting the LHC program with a \textit{neutrino-ion collider} in analogy with the planned electron-ion collider (EIC)~\cite{Accardi:2012qut}, which will operate at a similar center-of-mass energy and probe neutral current (NC) DIS.  
A high-statistics sample of neutrinos allows to measure their corresponding interaction cross sections differentially. Notably, the neutrino energies at the LHC significantly exceed those of any other previous experiments studying DIS neutrino interactions, such as CDHS~\cite{Berge:1987zw} or NuTeV~\cite{NuTeV:2005wsg}, and therefore provide an opportunity to extend their physics programs into a broader kinematic regime. Similar to the EIC and NuTeV program, collider neutrino measurements can be used to constrain proton and nuclear parton distribution functions (PDFs). The corresponding sensitivity was investigated in Ref.~\cite{Cruz-Martinez:2023sdv}, and it was found that high-statistics collider neutrino scattering measurement could significantly reduce PDF uncertainties. In particular, high resolution experiments with charm hadron identification capabilities provide an opportunity to measure the strange quark PDF via the charm associated neutrino interactions $\nu_\mu s \to \mu c$, while the typically heavy target nuclei permit to probe nuclear PDFs and constrain shadowing, anti-shadowing and the EMC effect for neutrinos. Ref.~\cite{Cruz-Martinez:2023sdv} also indicated that these measurements would then help to reduce PDF uncertainties for key measurements at the LHC central detectors, such as Higgs or weak boson production, and allow to break degeneracies between PDFs and possible new physics when interpreting LHC data~\cite{Hammou:2024xuj}.

The observed event rate depends on both the neutrino interaction cross section and the flux, the latter currently carrying significant uncertainties. While these uncertainties reduce sensitivity, several observables can help disentangle flux and interaction effects. For instance, the lepton momentum can reconstruct kinematic variables of DIS which are sensitive to partonic structure. Additionally, a comparison of event rates at different nuclear targets provides access to nuclear effects, while the fraction of events in exclusive channels like $\nu s \to \mu c$ probe specific initial quark flavors. Furthermore, channels with well-known cross sections can act as standard candles for flux measurements~\cite{Wilkinson:2023vvu}.\smallskip

\noindent \textbf{Neutrino Flux Measurements:} The neutrino beam at the LHC is produced via the decays of the lightest hadrons of a given flavor, most importantly charged pions, kaons and charm mesons. Notably, the production of these particles in the forward direction at LHC energies has not been measured before. Therefore, neutrino flux measurements provide a novel method to probe and constrain forward hadron production and provide insights into the underlying physics that cannot be obtained otherwise.  

Forward light-hadron production falls outside the scope of perturbative QCD. Instead, their production is described by phenomenological hadronic interaction models that were designed to match the available data~\cite{Pierog:2013ria, Ostapchenko:2010vb, Riehn:2019jet, Fieg:2023kld}. Previously, forward particle measurements at the LHC were restricted to neutral pions and neutrons, as obtained by the LHCf experiment~\cite{LHCf:2017fnw, LHCf:2018gbv}. Collider neutrino measurements will add complementary data on the forward production of charged pions, charged kaons and neutral kaons. Further constraining forward particle production at high energies and improving hadronic interaction models is particularly relevant for applications in astroparticle physics~\cite{FASER:2021pkt}, where these tools are used to describe astroparticle production in extreme astrophysical systems as well as cosmic ray interactions in the earths atmosphere. Notably, for the latter, there is a long-standing tension between the number of muons observed in high-energy cosmic ray air showers and the number predicted by hadronic interaction models. This discrepancy is often referred to as the \textit{muon puzzle}~\cite{PierreAuger:2014ucz, PierreAuger:2016nfk, EAS-MSU:2019kmv, Soldin:2021wyv, PierreAuger:2024neu}. This problem currently prevents cosmic ray observatories from inferring the mass composition of the cosmic ray flux and ultimately distinguish different hypotheses on their origins. Extensive studies over the last decade found that this discrepancy is most likely caused by a mismodeling of soft QCD effects in forward particle production at center-of-mass energies above the TeV scale~\cite{Ulrich:2010rg, Albrecht:2021cxw}. Further studies have shown that this problem could be resolved by suppressing neutral pion production via an enhanced rate of forward  strangeness production~\cite{Allen:2013hfa, Anchordoqui:2016oxy, Anchordoqui:2019laz}. This effectively increases the fraction of energy in the hadronic shower, and therefore the muon rate. Since strange hadron decays are one of the main sources of neutrinos at the LHC, measurements of collider neutrino fluxes allow to probe these scenarios and thereby help to resolve the muon puzzle. Indeed, studies using a phenomenological toy mechanism indicate that enhanced strangeness production could lead to substantial modifications of the LHC neutrino flux, which could be tested already with first data~\cite{Anchordoqui:2022fpn, Sciutto:2023zuz}.

In contrast to light hadrons, forward charm production can, in principle, be modeled using perturbative QCD methods~\cite{Bai:2020ukz, Maciula:2022lzk, Bhattacharya:2023zei, Buonocore:2023kna}. This contribution is particularly important for the production of high-energy electron neutrinos as well as tau neutrinos. Charm quarks are mainly produced via the gluon fusion process $gg \to cc$. A simple kinematic estimate shows that, in order to obtain TeV-energy neutrinos from charm decays, one of the initial state gluons needs to a carry a large momentum fraction $x \sim E_\nu/E_\text{proton} \sim 1$ while the other carries a very small momentum fraction $x \sim 4 m_c^2 / s \sim 10^{-7}$. The former regime would allow to constrain high-$x$ PDFs, complementing measurements performed at low-energy beam dump experiments, and may also shed light on intrinsic charm~\cite{FASER:2019dxq, Maciula:2022lzk}. The latter regime is sensitive to the gluon PDF at very low-$x$, well beyond the coverage of other experiments~\cite{Anchordoqui:2021ghd, Bai:2021ira, Abbaneo:2909524}. It therefore allows to study novel QCD phenomena, including BFKL dynamics and the onset of gluon saturation~\cite{Bhattacharya:2023zei}. Forward charm measurements at the LHC and a resulting better understanding of the underlying physics will also provide useful input for astroparticle physics. An example is the prompt atmospheric neutrino flux, which originates from the decay of charmed hadrons produced in cosmic ray collisions with the atmosphere. At the neutrino energies above a few 100~TeV, this process is expected to be the main background to searches for astrophysical neutrinos by neutrino telescopes such as IceCube~\cite{Abbasi:2021qfz}. However, there currently exist large uncertainties on the magnitude of the prompt atmospherics neutrino flux which limit the measurement of astrophysical parameters. Measurements of forward charm production at the LHC will help to reduce these uncertainties and improve the sensitivity for astrophysical neutrino measurements as well as multi-messenger astronomy~\cite{Bai:2022xad}. \smallskip

\noindent \textbf{Searches for BSM Physics:} The high energy of the neutrino beam and the potentially large number of interactions make collider neutrino experiments also an interesting laboratory for searches for new physics in the neutrino sector. A variety of signatures and scenarios have been investigated in the literature. This includes searches for modulinos or sterile neutrinos with multi-eV masses, leading to visible oscillation patterns~\cite{FASER:2019dxq, Anchordoqui:2021ghd, Bai:2020ukz, Anchordoqui:2023qxv, Anchordoqui:2024ynb}; searches for new neutrino-philic mediators that modify the predicted tau-neutrino flux~\cite{Kling:2020iar, Batell:2021snh}; searches for anomalous electromagnetic properties of neutrinos~\cite{MammenAbraham:2023psg} and transition dipole moments to heavy sterile neutrinos~\cite{Ismail:2021dyp};  searches for neutrino non-standard interactions that modify neutrino production or neutrino scattering~\cite{Ismail:2020yqc, Falkowski:2021bkq, Kling:2023tgr};  searches for neutrino self-interactions and neutrino-philic mediators to dark matter~\cite{Kelly:2021mcd, Berryman:2022hds}; and searches for BSM neutrino interactions through measurements of neutrino trident production~\cite{Francener:2024wul, Altmannshofer:2024hqd}.

Collider neutrino experiments, in particular SND@LHC, have also been proposed to search for the scattering of other (typically light) particles that can be abundantly produced via hadron decays in the forward direction~\cite{Boyarsky:2021moj}. This includes, for example, light dark matter and other light dark sector particles that may scatter on either electrons or nucleons in the detectors~\cite{Batell:2021blf, Batell:2021aja, Kling:2022ykt}. Unlike the neutrino signal, these typically lead to sub-GeV energy deposits in the detector that need to be identified. Another example are the ultra-light axions, which could convert into an energetic photon inside the electromagnetic fields of the nuclear target~\cite{Kling:2022ehv}. The LHC neutrino detectors are also exposed to a large flux of highly energetic muons, thereby effectively providing a muon fixed target experiment. This setup could be used to search for muonic force carriers~\cite{Ariga:2023fjg} or signs of lepton flavor violation~\cite{Batell:2024cdl}. Finally, in addition to scattering signatures, the  FASER experiment are also designed to look for decays of a broad array of proposed new particles~\cite{FASER:2018eoc} and first results were already obtained~\cite{FASER:2023tle, FASER:2024bbl}.

\section{EXPERIMENTS} 

\begin{figure}[hb]
\centering
\includegraphics[width=1.\textwidth]{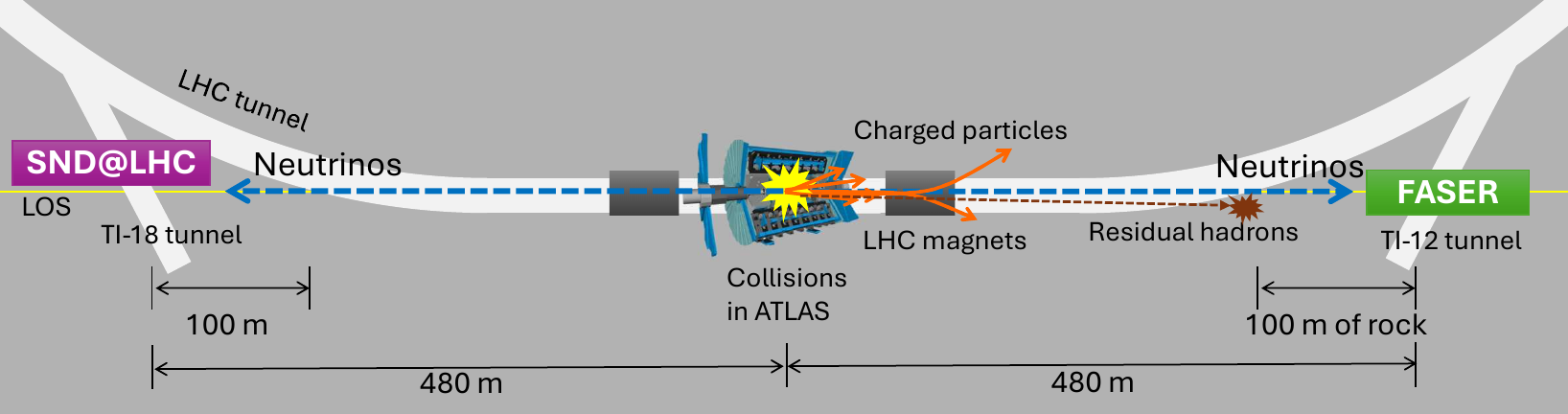}
\caption{\textbf{Sketch showing the location of the FASER and SND@LHC experiments with respect to the ATLAS collision point in the LHC.} (Figure not to scale).
}
\label{fig:location}
\end{figure}

The FASER and SND@LHC experiments are situated symmetrically, about 480-m from the ATLAS interaction point, and are close to aligned with the collisions-axis line of sight (LOS). As shown in Figure~\ref{fig:location}, both experiments are located in unused side tunnels that were formerly used to inject electrons/positrons into the LEP collider, but which are obsolete for the LHC. These locations are well suited for neutrino experiments at the LHC, since they allow the detectors to be placed close to where the neutrino flux, and energy, is maximum and the experiments are well shielded from the collision backgrounds. Particles produced in the LHC collisions need to traverse strong LHC magnets (which sweep away charged particles) and 100~m of concrete and rock before reaching the detectors. This leaves only a background from high-energy muons, with a flux of about 0.5~Hz cm$^{-2}$. The muon flux has been measured by the experiments~\cite{FASER:2018bac,SNDLHC:2023mib} and is also simulated using FLUKA~\cite{Battistoni:2015epi}.

The LHC beams are collided with a small crossing angle, to avoid additional collisions away from the IP. This has the effect of moving the collisions axis by around 7.5~cm at the experiment locations, with the crossing angle direction (up or down or horizontally away from the LHC machine) occasionally changed during LHC operations. The crossing angle choice, therefore changes the pseudorapidity~\footnote{Pseudorapidity is defined as $\eta \equiv -\log(\tan(\theta/2))$, with $\theta$ the polar angle with respect to the beam line. For neutrinos rapidity and pseudorapidity are equivalent, since neutrinos are essentially massless.} coverage of the experiments and the expected neutrino interaction rates. 

Both the FASER and SND@LHC experiments utilize a hybrid detector configuration for neutrino detection, combining information from passive nuclear-emulsion detectors with information from electronic detector components. The emulsion films are interleaved between tungsten plates to form an emulsion cloud chamber (ECC)~\cite{Ariga:2020lbq}, which acts as an instrumented neutrino target. The emulsion film records signals from all charged particles traversing the detector while it is in place, and therefore needs to be replaced periodically to keep the hit multiplicity low enough for track reconstruction and physics analysis.  Both experiments therefore need to exchange the emulsion detector after 20 - 30~fb$^{-1}$ of data, equivalent to a track density of about $5 \times 10^5$ tracks/cm$^2$. This corresponds to exchanging the emulsion films 3 to 4 times per year during LHC Run~3. 
The extremely precise charged particle tracking in the emulsion enables the detector to 
identify electrons from their electromagnetic shower, muons as long tracks traversing several interaction lengths of tungsten, and short-lived particles such as charm-hadrons and tau-leptons. This makes it very well suited to study all flavors of CC neutrino interactions, by reconstructing neutrino interaction vertices with an associated high-energy lepton.

\subsection{FASER}

\begin{figure}[hb]
\centering
\includegraphics[width=1.\textwidth, trim={0 15mm 0 20mm},clip]{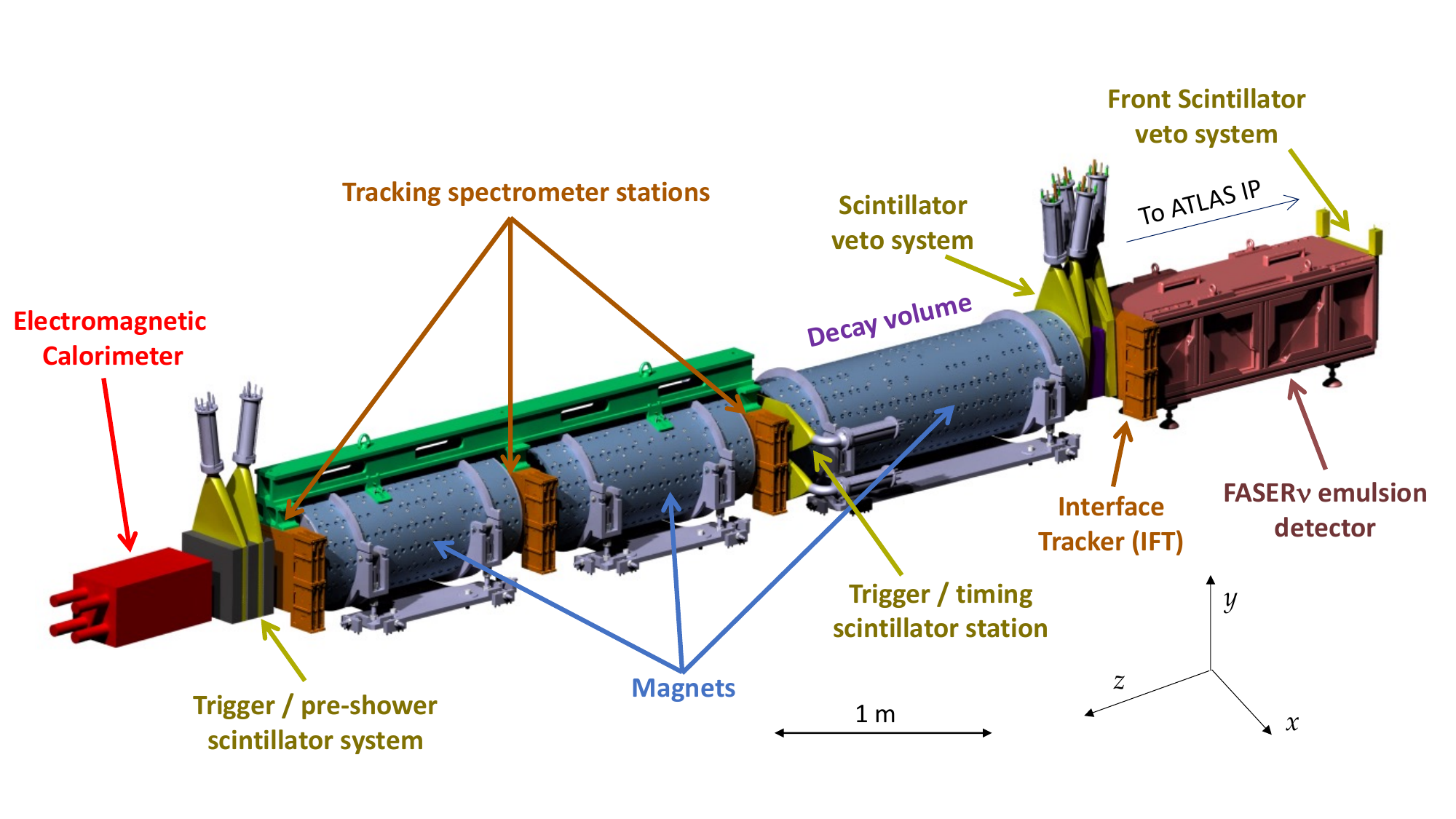}
\caption{\textbf{Schematic layout of the FASER detector.} The neutrino beam comes from the right (upstream) to the left (downstream). Figure adapted from Ref.~\cite{FASER:2022hcn} (CC BY 4.0). }
\label{fig:FASERdetectorsketch}
\end{figure}

The FASER detector is described in detail in Ref.~\cite{FASER:2022hcn}, and a sketch is shown in Figure~\ref{fig:FASERdetectorsketch}. 
The detector is aligned to the collision axis within a few cm, and for the crossing angle configuration in 2022/23 corresponded to a pseudorapidity coverage of $\eta > 8.8$. Neutrinos can be studied in FASER using both the dedicated FASER$\nu$ tungsten-emulsion sub-detector, and using the electronic detector components~\cite{Arakawa:2022rmp}. 

FASER$\nu$ is situated at the upstream end of FASER. It is made up of 730 $\times$ 1.1~mm-thick tungsten plates, interleaved with nuclear-emulsion films~\cite{Ariga:2020lbq}, with a transverse size of 30~cm $\times$ 25~cm. The target mass is 1.1~tonnes, corresponding to 220 radiation lengths and 7.8 hadronic interaction lengths. The emulsion films are each composed of 65~$\mu$m-thick emulsion layers on either side of a 210~$\mu$m-thick plastic base.  
To exploit the extremely precise position resolution for charged-particle tracks in the emulsion films, the alignment between the films needs to be stable over the exposure period. This is ensured by vacuum packing modules of 10 films with their corresponding tungsten plates, and then applying a strong pressure between the 73 installed modules. The temperature of the FASER$\nu$ detector is also kept constant at the 0.1 degree level at around 16.5 degrees by a dedicated cooling system.

Once the emulsion detector is removed from FASER it is disassembled and the emulsion films undergo a chemical development in the dark room at CERN, before being shipped to Japan where they are scanned with a high resolution scanning microscope~\cite{Yoshimoto:2017ufm} and the digitized data is then reconstructed. After a multistage alignment procedure with high energy background muons, the position resolution of track segments is found to be 300~nm in each projection~\cite{FASER:2024hoe}. 

The FASER electronic detector is made up of several veto scintillators, a tracking spectrometer~\cite{FASER:Tracker}, and an electromagnetic calorimeter. The spectrometer consists of three tracking stations (each made up of three layers of double-sided silicon microstrip tracking detector modules) interleaved with two 0.57~T 1~m-long dipole magnets. The magnet aperture defines the 20~cm-diameter acceptance of the detector in the transverse plane. Upstream of the spectrometer is a similar 1.5~m-long magnet which surrounds the decay volume for BSM particle searches. Upstream of this is a scintillator veto station and another tracking station which is designed to allow the combination of the FASER$\nu$ data with that from the electronic detector. At the front of the detector sits the FASER$\nu$ detector with two additional veto scintillators situated at the upstream end of this. The detector is triggered by signals in any of the scintillators or an energy deposit in the calorimeter of above about 20~GeV, leading to a typical trigger rate of 1.5~kHz, with a detector dead time of around 2\%~\cite{FASERTDAQ:2021}.

The FASER tracker has a position resolution of better than 30~$\mu$m in the precision plane (which corresponds to the bending plane of the magnets), and the calorimeter energy resolution is around 1\% for high energy electromagnetic deposits, as measured in a test beam.

\subsection{SND@LHC}

\begin{figure}[thb]
\centering
\includegraphics[width=1.\textwidth, trim={6mm 6mm 6mm 6mm},clip]{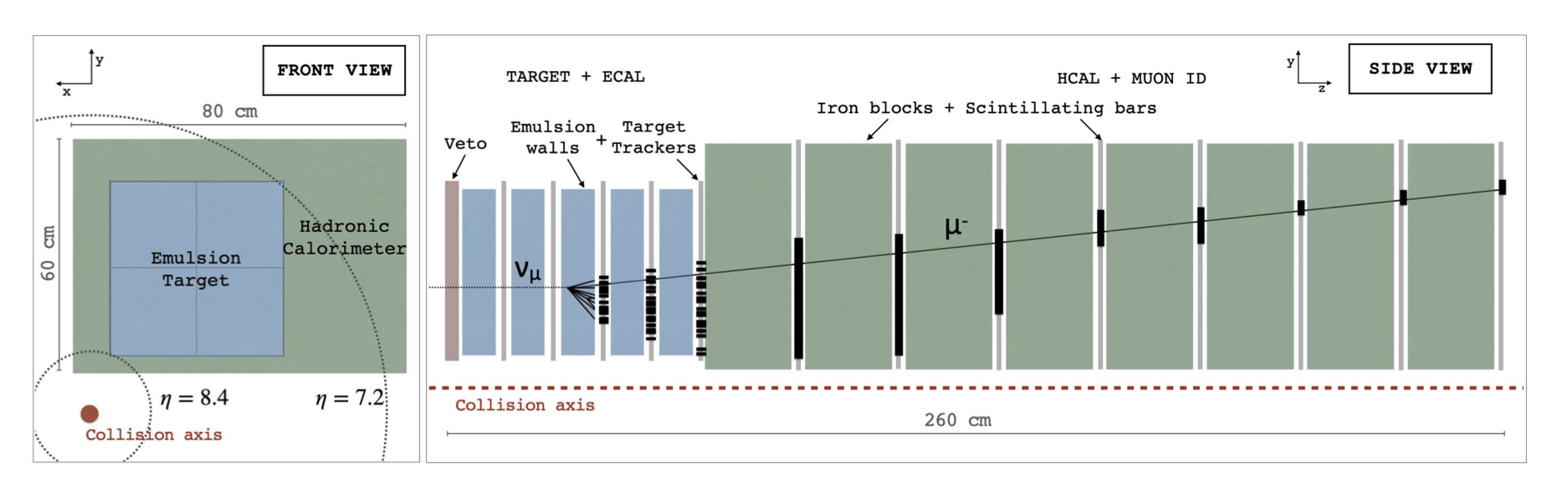}
\caption{\textbf{Schematic layout of the SND@LHC detector.} The pseudorapidity $\eta$ values are the limits for particles hitting the lower left and the upper right corner of the ECC. The side view includes an illustration of a simulated $\nu_\mu$ CC interaction. In the side view, the neutrino beam comes from the left (upstream) to the right (downstream). Figure reproduced from Ref.~\cite{SNDLHC:2023pun} (CC BY 4.0). }
\label{fig:SND@LHCdetectorsketch}
\end{figure}

The SND@LHC experiment is primarly optimized to detect and measure neutrino interactions, from neutrinos produced in decays of particles produced in collisions at IP1. The experiment has however also sensitivity to a number of new particles as expected to be produced in a number models beyond the Standard Model~\cite{Boyarsky:2021moj}.

The SND@LHC experiment was first proposed in 2020~\cite{SHiP:2020sos}, and was capitalizing on the gained experience in designing a neutrino detector for SHiP~\cite{SHiP:2015vad}, an experiment proposed to operate at the CERN  400 GeV SPS proton beam. The continuing synergy between SND@LHC and SHiP is of vital importance for both detectors. SND@LHC was approved at CERN in March 2021, and in November the same year the installation of the detector started in the foreseen tunnel. The installation was concluded before the start of Run 3 of the LHC in 2022; the detector was ready and partially commissioned just in time for the first physics data taking period of that run.
The SND@LHC detector is described in detail in Ref.~\cite{SNDLHC:2022ihg} and a sketch is shown in Figure~\ref{fig:SND@LHCdetectorsketch}. 
 
The electronic detectors provide the time stamp of the neutrino interaction and pre-select the interaction region while the neutrino interaction vertex is reconstructed using tracks in the tracking detectors and the emulsion. The upstream veto system is used to tag background muons and other charged particles entering the detector from the IP1 direction.
 
The veto system consists of two parallel planes of scintillating bars. Each plane is made of seven $1 \times 6 \times 42$~cm$^3$ vertically stacked bars of plastic scintillator. The target section with a mass of about 800 kg contains five walls. Each wall consists of four units,
called bricks, of ECCs
and is followed by a scintillating fiber (SciFi) station for tracking.
Each ECC unit is a sequence of 60 nuclear emulsion films of 19.2 $\times$ 19.2 cm$^2$ and approximately 
300 $\mu$m thick, interleaved with 59 tungsten plates, each 1 mm thick.
Details on the emulsion treatment are similar to what is reported in the previous section for FASER. Each SciFi station consists of one horizontal and one vertical $39 \times 39$~cm$^2$ plane. Each plane comprises six staggered layers of 250~$\mu$m diameter polystyrene-based scintillating fibers. The single particle spatial resolution in one plane is roughly 100~$\mu$m and the time resolution for a particle crossing both $x$ and $y$ planes is about 250~ps.

The muon system consists of two parts: the first five stations (UpStream, US), and the last three stations (DownStream, DS). Each US station consists of 10 stacked horizontal scintillator bars of $82.5 \times 6 \times 1$~cm$^3$, resulting in a coarse $y$ view and together they act mainly as a hadronic calorimeter with a depth of about 9.5 hadronic interaction lengths. 

A DS station consists of two layers of thinner bars measuring $82.5 \times 1 \times 1$~cm$^3$, arranged in alternating $x$ and $y$ planes, leading to a spatial resolution in each axis of less than 1 cm for muon reconstruction. The eight scintillator stations are interleaved with 20~cm thick iron blocks. Events with hits in the DS detector and the SciFi tracker are used to identify muons.

All signals exceeding preset thresholds are read out by the front-end electronics and clustered in time to form events. An efficient software noise filter is applied to the events online, resulting in negligible detector deadtime and negligible loss in signal efficiency. Events satisfying certain topological criteria, such as the presence of hits in several detector planes, are read out. In the absence of beam, the noise filtering logic reduces the event rate by five orders of magnitude to around 4~Hz. At the highest instantaneous luminosity in 2022 and 2023 ($2.5 \times 10^{34}$~cm$^{-1}$~s$^{-1}$), this set up generated a 
data rate of around 5.4~kHz.

\subsection{Comparison between the experiments}
\label{sec:expComp}

\begin{table} 
\setlength{\tabcolsep}{4pt}
\begin{center}
 \begin{tabular}{c c c c }
\hline
\hline
\ Experiment &  \ Pseudorapidity Coverage & \ Target  & \ Detector \\  
 \hline
\ FASER &  $\eta > 8.8 $ & 1100 kg  & Emulsion vertex detector \\
& & tungsten & Tracking spectrometer \\
& &  & EM calorimeter \\
\hline
\ SND@LHC &  $7.2< \eta < 8.4 $ & 800 kg  & Emulsion and SciFi vertex/tracking detector \\
& & tungsten & EM calorimeter \\
& & & Hadronic calorimeter and muon system \\
\hline
\hline
\end{tabular}
\end{center}
\caption{Comparison between the most relevant features of the two experiments.}
\label{tab:expComp}  
\end{table} 

\begin{table} 
\setlength{\tabcolsep}{15.7pt}
\begin{center}
 \begin{tabular}{ c c c c c }
\hline
\hline
\ Experiment & Quantity & \ $\nu_e$ & \ $\nu_\mu$  & \ $\nu_\tau$ \\  
 \hline
FASER & N$_{\rm{int}}$ &  2331$^{+1227}_{-544}$ &  12014$^{+1145}_{-1636}$ & 46$^{+77}_{-21}$ \\ 
& average energy  & 785 GeV & 716 GeV & 849 GeV \\
\hline
SND@LHC &  N$_{\rm{int}}$ & 307$^{+307}_{-116}$& 1694$^{+297}_{-549}$ & 15$^{+26}_{-7}$\\
 &  average energy & 442 GeV & 357 GeV & 596 GeV \\
\hline
\hline
\end{tabular}
\end{center}
\caption{Expected number of neutrino interactions and the average interacting neutrino energy at FASER and SND@LHC during LHC Run~3 with 350~fb$^{-1}$ of data for the three flavors of neutrinos (numbers are for neutrinos and antineutrinos combined). No selection efficiencies were applied.}
\label{tab:expInt}  
\end{table} 

Table~\ref{tab:expComp} shows a comparison between the most relevant features of the two experiments in terms of their neutrino physics program. The two experiments are situated to cover slightly different pseudorapidity regions, which enables complementary physics studies. The FASER detector is on-axis, covering higher pseudorapidites, and therefore is exposed to a larger neutrino flux and higher neutrino energies. SND@LHC is situated slightly off-axis, corresponding to a lower pseudorapidity range, which means a higher fraction of the neutrinos observed by the experiment originate from charm-hadron decay.  

The resulting number of neutrino interactions expected within FASER and SND@LHC during LHC Run~3 with collisions at $\sqrt{s}$=13.6~TeV and an assumed luminosity of 350~fb$^{-1}$ is shown in Table~\ref{tab:expInt}. Also included are the average energies of the interacting neutrinos. These results were obtained using the simulation described in Ref.~\cite{Kling:2021gos}. The fast neutrino flux simulation introduced in Ref.~\cite{Kling:2021gos} with the LHC Run~3 configuration as described in Ref.~\cite{FASER:2024ykc} was used to model the propagation of light long-lived mesons though the LHC's beam pipe and magnetic fields as well as to simulate their decays. Following the recommendation of Ref.~\cite{FASER:2024ykc}, this prediction uses \texttt{EPOS-LHC}~\cite{Pierog:2013ria} to simulate the production of forward light hadrons and the envelope formed by \texttt{EPOS-LHC}, \texttt{SIBYLL~2.3d}~\cite{Riehn:2019jet}, \texttt{QGSJET~2.04}~\cite{Ostapchenko:2010vb} as well as the forward physics tune of \texttt{PYTHIA}~\cite{Fieg:2023kld} to define an uncertainty band. Using the results of Ref.~\cite{Buonocore:2023kna}, charm hadron production is modeled using \texttt{POWHEG}~\cite{Nason:2004rx, Frixione:2007vw, Alioli:2010xd} matched with \texttt{PYTHIA~8.3}~\cite{Bierlich:2022pfr} for parton shower and hadronization, with the uncertainties described by scale variations. The expected event rates were estimated using the neutrino interaction cross section provided by \texttt{GENIE}~\cite{Andreopoulos:2009rq}. The Bodek-Yang model~\cite{Bodek:2002vp, Bodek:2004pc, Bodek:2010km} employed in \texttt{GENIE} shows good agreement with more recent cross section calculations for high-energy neutrinos~\cite{Candido:2023utz, Jeong:2023hwe} within an uncertainty of $\lesssim 6\%$~\cite{FASER:2024ykc}. 

\begin{figure}[hb]
\centering
\includegraphics[width=1.\textwidth]{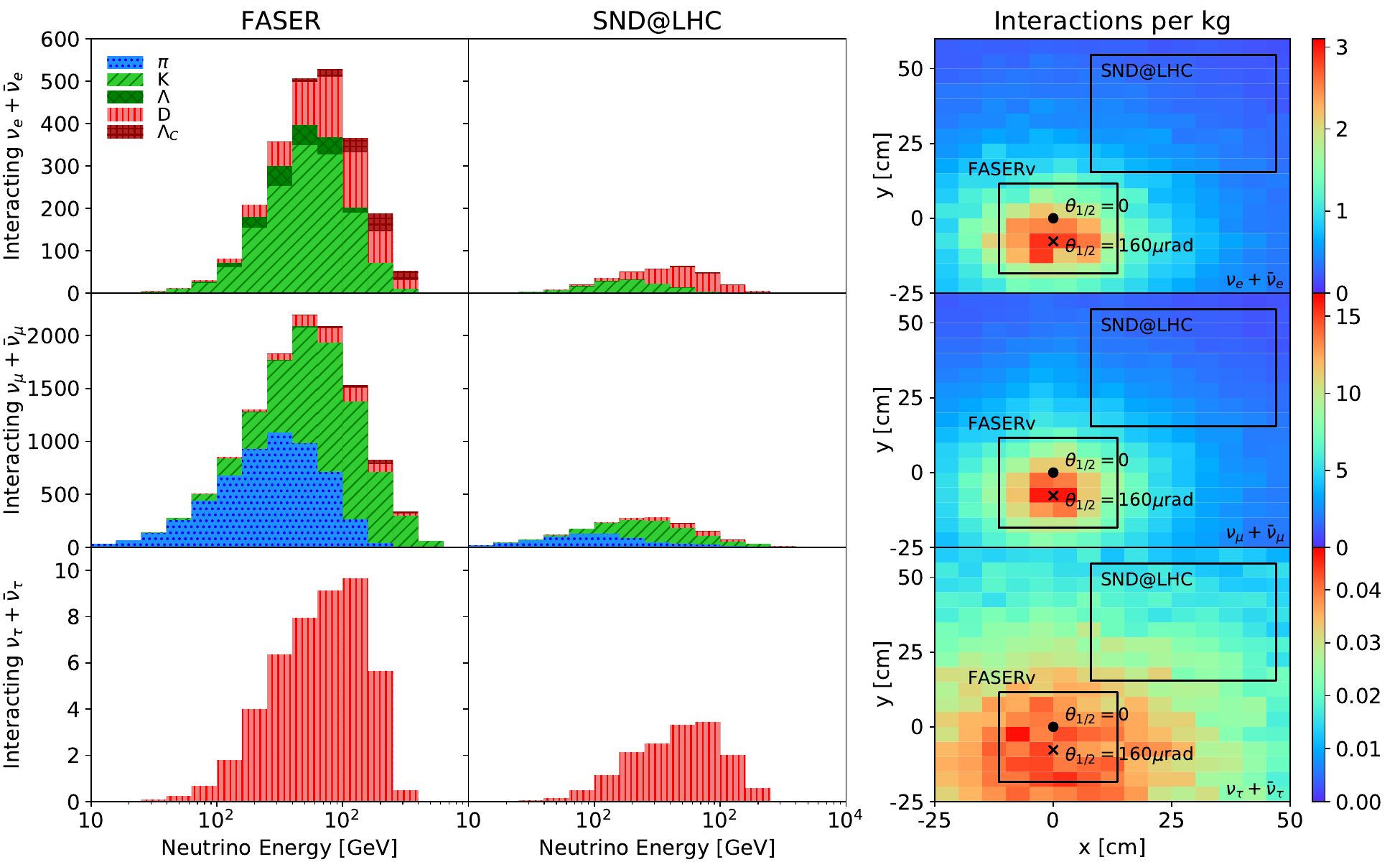}
\caption{\textbf{Energy and spatial distribution of interacting neutrinos.} The left and central panel show the expected number of neutrino interactions as function of the neutrino energy in the FASER$\nu$ and SND@LHC detectors for a luminosity of 350~fb$^{-1}$. The colored histograms indicate the flux composition in terms of the parent hadrons. The right panel shows the transverse distribution of interacting neutrinos at 480~m from the IP. The markers show the position of the beam collision axis without a beam crossing angle $\theta_{1/2}=0$ and for a half crossing angle of $\theta_{1/2}=160~\mu$rad downwards as in 2022 and 2023. The boxes indicate the transverse coverage of the two detectors. }
\label{fig:fluxes}
\end{figure}

Figure~\ref{fig:fluxes} shows the expected energy and transverse spatial distribution for interacting neutrinos in both detectors. Looking at the spatial distribution, one can see that the expected number of neutrino interactions per unit mass falls off quickly when moving away from the beam collision axis. This is largely due to the collimation of the beam, whose angular size is determined by the parent particle mass $ \theta = p_{T_\nu} / E_{\nu} \sim m_\text{parent}/ E_{\nu}$. The beam is more collimated for electron and muon neutrinos, which primarily originate from light hadron decay. Instead, tau neutrinos originate only from heavy hadron decays, leading to typically larger transverse momenta and therefore a broader beam. The neutrino energy also decreases away from the beam collision axis, which additionally reduces the rate of interacting neutrinos. Due to its off-axis location, less events are expected to occur in the SND@LHC detector compared to FASER$\nu$. 

The colored histograms in the left panels indicate the flux composition. Charged pion decays only contribute to the flux of muon neutrinos, where they provide the dominant contribution at lower energies below a few 100~GeV. Kaon decays contribute to both the muon neutrinos, providing the dominant contribution at higher energies, as well as electron neutrinos. Neutrinos from charm meson decays provide the leading contribution to the flux of high energy electron neutrinos. Their relative contribution to the flux increases when moving away from the beam collision axis, resulting in a larger fraction of neutrinos from charm hadron decays in SND@LHC compared to FASER. Decays of $\Lambda$ and $\Lambda_c$ baryons mainly contribute to the electron neutrino spectrum at FASER. Tau neutrinos are almost exclusively produced in charm-meson decays, with only a small additional contribution expected from beauty hadrons~\cite{Buonocore:2023kna}. 

\section{FIRST NEUTRINO OBSERVATION}
\label{sec:first-results}

The first neutrino interaction candidates were reported with data taken by FASER's pilot emulsion detector in 2018~\cite{FASER:2021mtu}, however, the reported significance of 2.7 standard deviations ($\sigma$) was not sufficient to establish an observation of neutrinos at the LHC. 

Following this, the FASER~\cite{FASER:2023zcr} and SND@LHC~\cite{SNDLHC:2023pun} experiments announced the first observation of neutrino interactions at the LHC in March 2023. Both experiments used the full 2022 dataset corresponding to about 36~fb$^{-1}$ of collision data. Since it takes time to process the emulsion films, both experiments used only information from the electronic detectors to search for muon neutrino CC interactions in the target. This leaves a clear detector signature of no signal in the scintillator veto systems upstream of the target, and then a high-energy muon track traversing the detector downstream of the target, with possible additional activity after the target from the hadronic recoil. The main backgrounds arise from inefficiencies of the veto system, background muons that miss the veto system, as well as neutral hadrons produced in muon interactions upstream of the detector, that could mimic the signal signature. Both experiments adopted a blind analysis technique in order to avoid unconscious bias.

\noindent \textbf{Observation at FASER:} In FASER, the analysis required a reconstructed charged-particle track in the spectrometer with momentum greater than 100~GeV, and that the extrapolated track must pass through the central region of the veto scintillators. About 130 signal neutrino events were expected to pass this selection. Each of the two veto scintillators' inefficiency was extracted from the data by measuring the rate of events where only one of the two scintillators had a signal. This gave a combined inefficiency of $\mathcal{O}$($10^{-15}$) and a negligible background from the veto inefficiency. Other background arising from muons that miss the front veto system, but lead to the signal selection topology, or from neutral hadrons produced upstream of the detector, were considered but found to also be negligible. The total background estimate was 0.08 $\pm$ 1.83 events. 

\begin{figure}[thb]
\centering
\includegraphics[width=1.\textwidth, trim={0 30mm 0 0},clip]{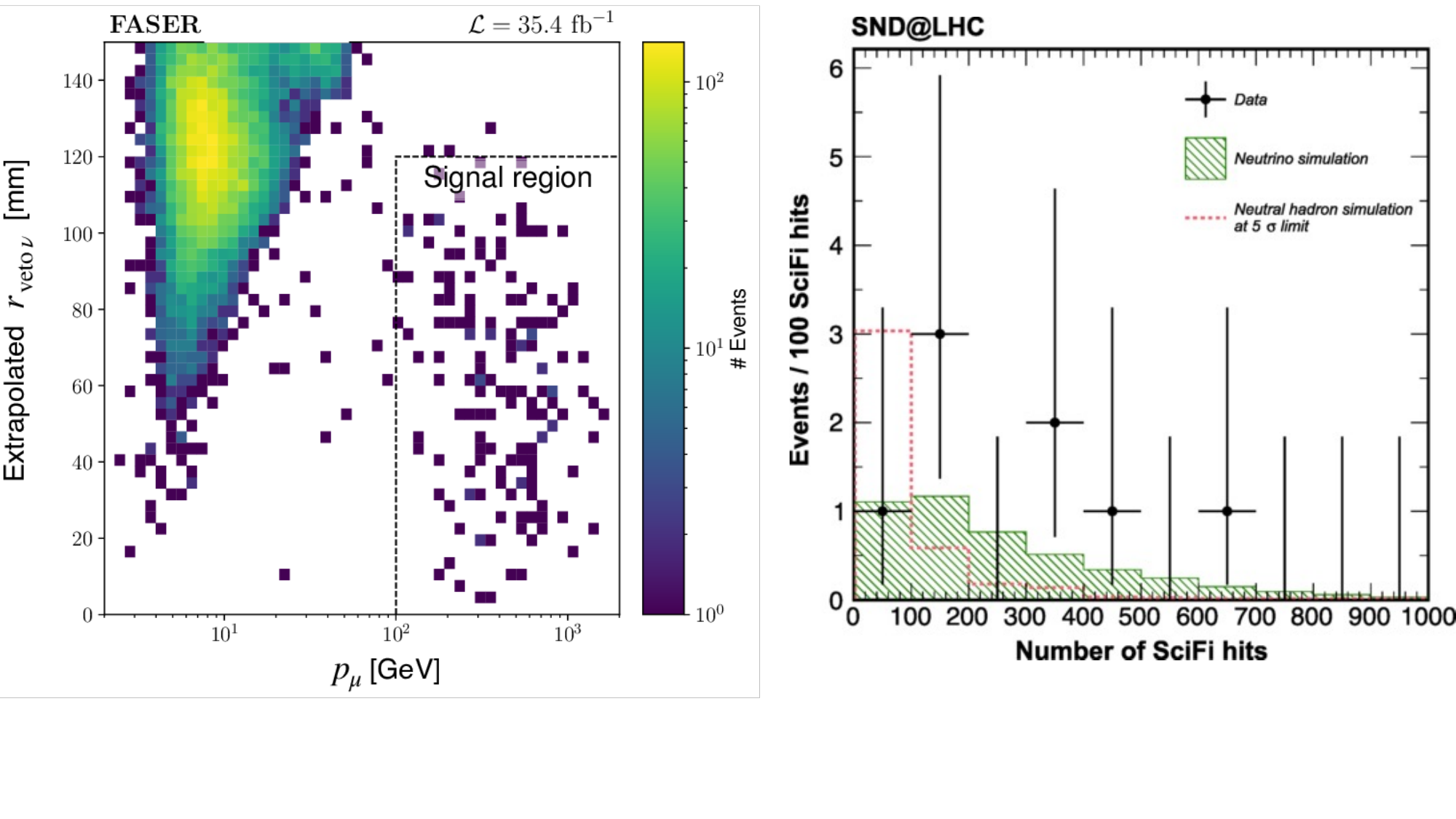}
\caption{\textbf{First observation of neutrinos at the LHC.} The left panel shows events selected by the FASER analysis. One can see a clear separation between the low momentum  background events and the high momentum neutrino signal. Panel reproduced from Ref.~\cite{FASER:2023zcr} (CC BY 4.0). The right panel presents the number of SciFi hits for the events selected by the SND@LHC analysis, showing an excess of events with large SciFi multiplicity compared to the background only hypothesis (red dashed line). The expected neutrino signal distribution is shown as a green hashed histogram. Panel reproduced from Ref.~\cite{SNDLHC:2023pun} (CC BY 4.0). } 
\label{fig:first-obs}
\end{figure}

Figure~\ref{fig:first-obs} (left), shows the selected events in the plane of the track momentum versus the track radius in the transverse plane when extrapolated to the front veto, relaxing the cuts on these variables, with the signal region shown as the black dashed square. The figure shows a clear separation between the background at low momentum, and with extrapolated tracks close to the edge of the veto system, and the neutrino signal events. 

The analysis selected 153 data events in the signal region, corresponding to a statistical significance of observing the neutrino signal at the level of 16$\sigma$. 
Examining the charge and momentum of the muon track in the selected events showed that both neutrinos and antineutrinos were observed, with the ratio broadly consistent with expectations. The sample contained events with track momentum above several hundred GeV, meaning the highest energy detected neutrino interactions from an artificial source were observed.

\noindent \textbf{Observation at SND@LHC:} In SND@LHC the analysis to observe $\nu_{\mu}$  CC interactions from LHC collisions, fully based on the electronic detectors, required a veto on a charged particle in the upstream veto system, a muon track in the muon system and a shower in the target area as observed in the SciFi tracker and HCAL detector.
The data sample was collected during the first year corresponding to a total integrated luminosity of 36.8~fb$^{-1}$ and was collected with an efficiency of 95\%.

The expected neutrino signal and background event production in proton-proton collisions at the LHC was simulated with Monte Carlo programs and the background validated by measurements with data.  
A total of around $1.6 \times 10^5$ simulated neutrino events and $3 \times 10^7$ background events were generated and processed.

Considering the mass of the tungsten target during the 2022 run (approx. 800 kg), about $157\pm 37~\nu_{\mu}$ CC DIS interactions are expected in the full target of the analyzed data set. The large range in the expectation is caused mostly by the difference between the predictions of the $\nu_{\mu}$ flux at SND@LHC from different models used as reported in Ref.\cite{Kling:2021gos}. The modeling is complex, and the different Monte Carlo programs have associated uncertainties ranging from 10 to 200\%. Observing the rare neutrino signal over the prevailing background implies adopting a selection with strong rejection power and restricted fiducial volume designed to yield a clean set of events. As a result of the full selection, 8~$\nu_{\mu}$ CC DIS candidates are identified, with 4.5 events expected~\cite{SNDLHC:2023pun}.  Figure~\ref{fig:first-obs} (right) shows the number of SciFi hits for selected events in data compared to MC predictions for signal and background.

To estimate the background from penetrating muons, the inefficiency of the veto system is determined from data. The overall veto system inefficiency during 2022 was $4.5 \times  10^{-4}$~\cite{SNDLHC:2024bzp}.  
The corresponding SciFi inefficiency per station was $1.1 \times 10^{-4}$, making the combined inefficiency of the veto system and the two most upstream SciFi planes $5.3 \times 10^{-12}$. 
The signal event rate is significantly lower in SND@LHC compared to FASER due to the very stringent selection cuts imposed on the data of this novel detector for this early analysis, the off-axis location of the experiment, and to avoid an observed veto inefficiency region.

The background yield after the selection amounts to $(7.6 \pm 3.1) \times 10^{-2}$ and is dominated by neutrons and $K_L^0$s. This results in a p-value of $1.48 \times 10^{-12}$, corresponding to an exclusion of the background-only hypothesis at the level of 7.0 standard deviations. Hence this result constitutes a clear direct observation of neutrinos produced in proton-proton collisions at the LHC.

\section{LATEST RESULTS}

\subsection{FASER$\nu$ emulsion detector results}

In March 2024, FASER released the first results using the FASER$\nu$ emulsion detector~\cite{FASER:2024hoe}. In this analysis, a small subset of the first emulsion detector exposed to 9.5~fb$^{-1}$ of data, corresponding to a target mass of 128.8~kg, was used to search for $\nu_e$ and $\nu_\mu$ CC interaction candidates. These candidates produce the signature of high-energy neutral vertices, defined as vertices with no associated upstream charged particle tracks, accompanied by high-energy electron ($e$) or muon ($\mu$) candidates, respectively. 
The background to such vertices arises from neutral hadrons (produced upstream by high energy background muons) interacting in the detector. However, the neutral hadrons are typically much lower energy and are not associated with a high-energy lepton ($e$ or $\mu$). The background is significantly reduced using topological and kinematic cuts to select high-energy vertices, and then reduced to a negligible level by the selection of an associated lepton with reconstructed energy of more than 200~GeV and by requesting the back-to-back topology of the lepton and hadron system. 
Electron candidates are reconstructed from the track segments produced in the electromagnetic shower in the detector, and the energy is estimated from the segment multiplicity at the shower maximum. Simulation studies show this gives an energy resolution of 25\% for electrons with energy of 200~GeV, and this has since been validated with test-beam data. Muon candidates are selected as long tracks penetrating more than 100 tungsten plates and the momentum is estimated using the effect on the track trajectory due to multiple Coulomb scattering in the detector~\cite{FASER:2019dxq}. The momentum resolution is 30\% at 200~GeV, which has also been validated with test-beam data \cite{fujimori_2024_13222911}.

The background from neutral hadron interactions is estimated from high statistics Monte Carlo simulation samples, where the simulation is validated by comparing data and simulation for low-energy neutral vertices without identified leptons which are dominated by neutral hadron interactions. For the $\nu_\mu$ selection there is also a background from NC neutrino interactions, which is estimated from simulation. The total expected background is 0.025$^{+0.015}_{-0.010}$ ( 0.22$^{+0.09}_{-0.07}$) events for the $\nu_e$ ($\nu_\mu$) channel. Four data events were selected by the $\nu_e$ selection, and eight by the $\nu_\mu$ selection, to be compared to the expectation of 1.1–3.3 events, and 6.5–12.4 events, respectively. The statistical significance of the observation is 5.2$\sigma$ for the observation of $\nu_e$ interactions and 5.7$\sigma$ for the observation of $\nu_\mu$ interactions. This therefore constitutes the first observation of electron neutrino interactions at a collider.

\begin{figure}[htb]
\centering
\includegraphics[width=\textwidth]{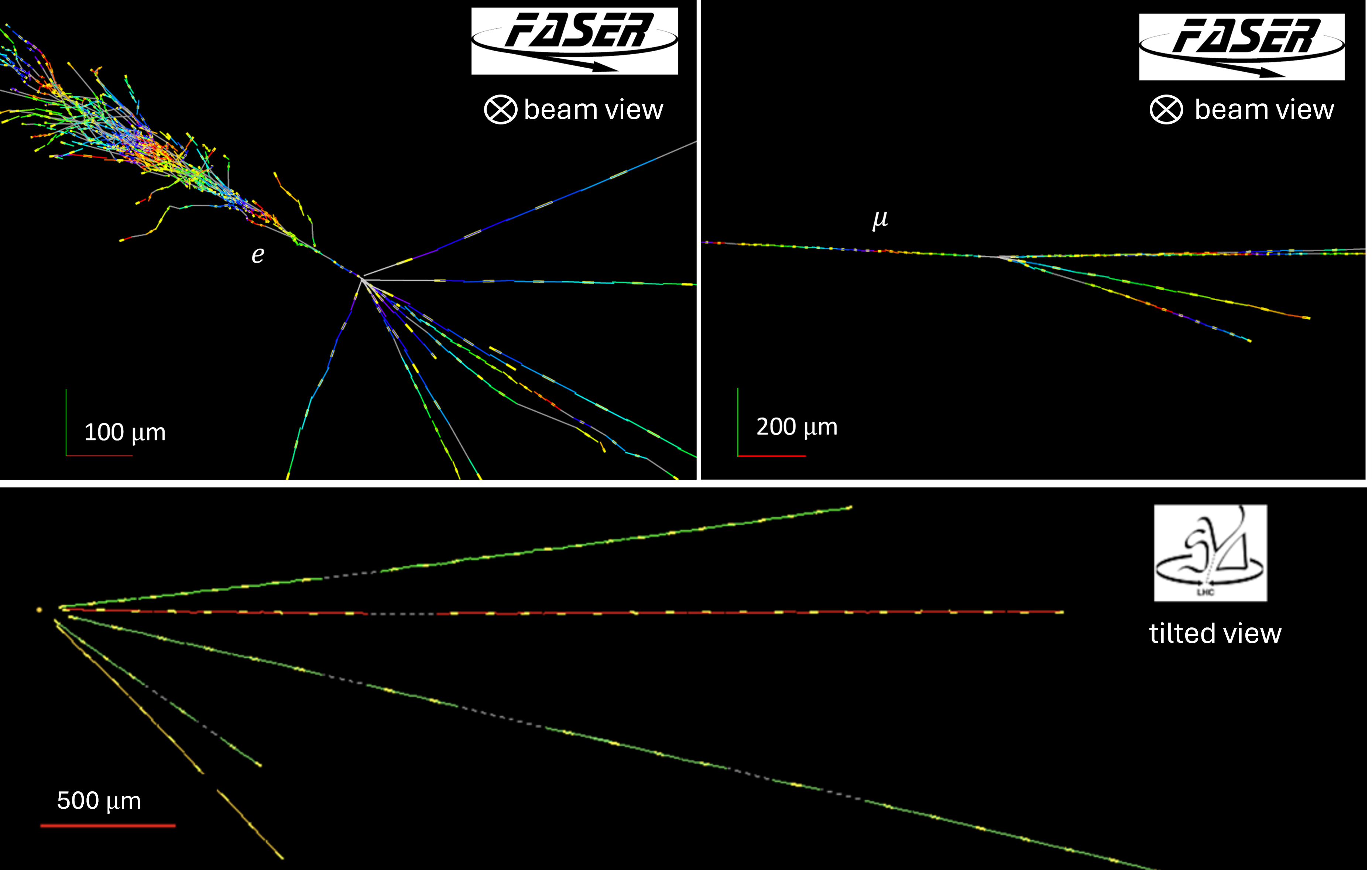}
\caption{\textbf{Example neutrino vertices.} The upper panel shows a $\nu_e$ (left) and a $\nu_\mu$ (right) candidate in a transverse plain with respect to the neutrino beam direction found in FASER$\nu$. The kinematics of leptons and hadrons show a back-to-back feature expected from CC neutrino interactions. Panel adapted from Ref.~\cite{FASER:2024hoe} (CC BY 4.0).
The bottom panel shows a tilted view of a vertex found in SND@LHC. }
\label{fig:FASERnu-cand}
\end{figure}

\begin{figure}[hb]
\centering
\includegraphics[width=1\textwidth]{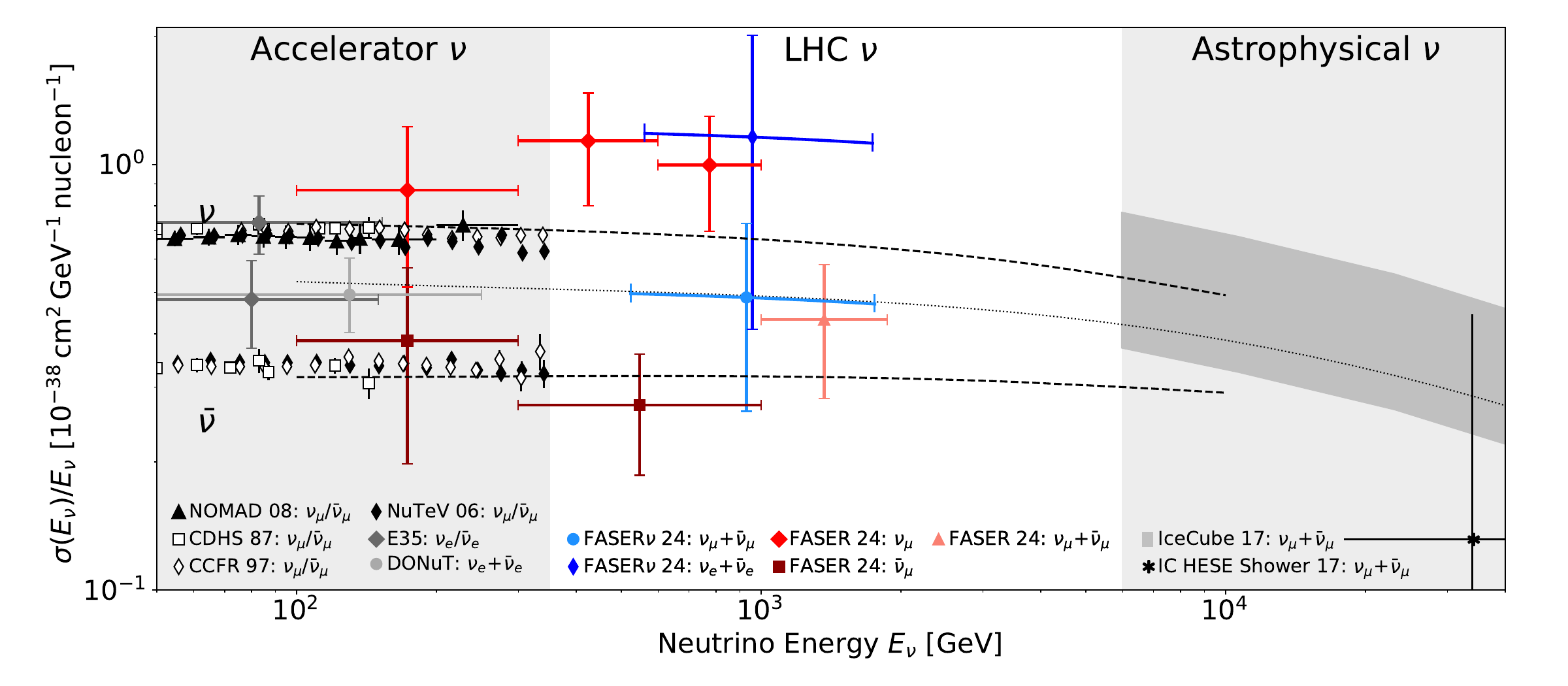}
\caption{\textbf{Neutrino cross section measurements.} The measured neutrino-nucleon cross sections, normalized to the neutrino energy, using  FASER$\nu$~\cite{FASER:2024hoe} and FASER~\cite{FASER:2024ref} compared to previous measurements from fixed-target~\cite{NOMAD:2007krq,Berge:1987zw,Seligman:1997fe,NuTeV:2005wsg,Baltay:1988au,DONuT:2007bsg} and astrophysical sources~\cite{IceCube:2017roe,Bustamante:2017xuy,IceCube:2020rnc}. The dashed contours correspond to the cross sections predicted by the Bodek-Yang model, as implemented in GENIE. Note that the displayed experiments do not all
use the same targets.}
\label{fig:FASERnu_xsec}
\end{figure}

These results were used to make the first measurement of the neutrino interaction cross section at TeV energies. The cross section is calculated by scaling the cross section predicted by the Bodek-Yang model with the ratio of the observed-to-expected number of events in each channel. Figure~\ref{fig:FASERnu-cand} (top panels) shows example selected neutrino interaction vertices, and Figure~\ref{fig:FASERnu_xsec} shows the measured cross sections (blue markers). Since the emulsion does not allow to measure the charge of the produced lepton, it is not possible to separate the neutrino and antineutrino events, so the measured cross section corresponds to the weighted average of these. In the figure the cross section is shown versus the neutrino energy in a single bin, which corresponds to the 68\% energy range of the expected interacting neutrino energy from simulation, and nicely shows the complementarity of the energy of LHC neutrinos measurements compared to those from fixed target and neutrino telescope experiments. The results agree with the SM expectation. It should be noted that this result uses only 1.1~\% of the data already collected by FASER$\nu$ (up to the end of the 2024 LHC run), so there will be a large increase in the statistics used for future results.

\subsection{Measurement of the energy dependent cross section and neutrino flux}

In December 2024, FASER released the first measurement of the muon neutrino interaction cross section in bins of neutrino energy and separated for neutrino and antineutrino~\cite{FASER:2024ref}. As for the first observation of neutrinos at the LHC, this analysis focused on muon neutrino CC interactions in the FASER$\nu$ target with the electronic detector components, and uses the measured momentum and charge of the reconstructed muon candidate to infer if the interaction was from a neutrino or antineutrino and its energy. The measurement was carried out using the 2022 and 2023 dataset corresponding to about 66~fb$^{-1}$ of data. Events are selected in a similar way to the previous analysis (discussed in Section~\ref{sec:first-results}) with some small modifications to improve the signal efficiency and to reduce the systematic uncertainties. 

The signal is defined as $\nu_\mu$ CC neutrino interactions in a fiducial volume corresponding to a 200~mm diameter cylinder in the target, centered on the electronic detector axis, and with neutrino energy greater than 100~GeV.  The analysis selects 362 data events, corresponding to $338.1 \pm 21.0$ observed $\nu_\mu$ CC candidates in the defined fiducial volume, with $298.4 \pm 42.6$ events expected. The background is dominated by non-$\nu_\mu$ CC neutrino interactions, and $\nu_\mu$ CC neutrino interactions outside the fiducial volume. Correcting for the selection efficiency yields a total of 1242.7 $\pm$ 137.1 CC muon neutrino interactions inside the fiducial volume.

For signal events passing the event selection, the muon retains on average 80\% of the neutrino energy. The {\it calibrated} muon momentum $p'_\mu$=$p_\mu$/0.8 is therefore used to determine the number of selected events as a function of neutrino energy, as follows. The selected events are divided into six bins of the ratio of charge to calibrated momentum ($q/p'_\mu$) with bin edges:
\begin{align}
\left[ -\frac{1}{100}, -\frac{1}{300}, -\frac{1}{600}, -\frac{1}{1000}, \frac{1}{1000}, \frac{1}{300}, \frac{1}{100} \right] \, \ \rm{/GeV} \, . \nonumber
\end{align} 
This gives three momentum bins for negatively charged muons (which originate from neutrino interactions), two momentum bins for positively charged muons (from antineutrino interactions), and one high momentum bin including muons of either charge with $p'_\mu >$ 1~TeV, since the charge identification performance degrades at high momentum. The binning is defined to have a similar number of expected events in each bin. The selected events are unfolded to give the number of (anti)neutrino events in bins of neutrino energy with the same bin ranges. The results are interpreted in two ways, firstly the expected neutrino flux (from simulation, as described in Section~\ref{sec:expComp}) is used to derive the (anti)neutrino cross section as a function of energy, which is shown in Figure~\ref{fig:FASERnu_xsec} (red markers). Secondly, the theoretical cross section is assumed and the neutrino flux is derived from the data. The measured flux is then used to put a first constraint on forward charged pion and kaon production at the LHC. The dominant uncertainties affecting the cross section measurement arise from the event statistics and systematic uncertainties related to the neutrino flux, and the modeling of the neutrino interactions, whereas experimental systematic uncertainties are sub-dominant.

\subsection{New SND@LHC results}

The most recent results from SND@LHC include further analyses of the 2022 data discussed before, combined with data collected during 2023. The overall veto system inefficiency during 2023 was $6.6 \times 10^{-6}$~\cite{SNDLHC:2024bzp}.  A total integrated luminosity of 
31.8 fb$^{1}$ was recorded in 2023, with an efficiency of 99.7 \%.
A slightly modified version of the first published data analysis to select a clean $\nu_{\mu}$ neutrino sample has been performed including the 2023 data, resulting in a total number of 32 selected events~\cite{SNDMoriond24} in the full data sample, leading to an observation significance of the $\nu_{\mu}$ signal now being well over 10 standard deviations.

The analysis on the emulsion data is ongoing in four scanning labs namely CERN, Naples, Bologna and Santiago de Chile using a total of 10 scanning microscopes. The scanning of emulsion film is very time intensive; about 60\% of the 2022 data had been scanned by November 2024. Calibrations and physics analyses have been started using these scanned samples. The aim is to achieve a 200~nm spatial resolution. 
An example of a neutrino candidate in the emulsion stack is shown in Figure~\ref{fig:FASERnu-cand} (bottom).

In addition, SND@LHC has performed an analysis which has sensitivity to other than muon neutrino flavor interactions, based solely on the electronic part of the detector~\cite{SNDLHC:2024qqb}. The analysis is based on selecting an interaction vertex in the neutrino target i.e. not associated with an incident charged particle, and on requiring that there is no muon track produced at the selected vertex, as well as to have large activity in the scintillating fiber detector in the target consistent with an electromagnetic or hadronic shower. 
These events are labeled here as $\nu0\mu$. Such signal events can be interpreted as originating from a mix of electron and tau neutrino charged current events, plus neutral current events. The background consists of muon neutrino charged current events where the muon escaped detection or was otherwise not reconstructed, and interactions of neutral hadrons in the neutrino target stemming from mostly muon DIS events produced upstream of the SND@LHC detector that penetrated into the  target. 
A detector variable that measures the activity in the neutrino target, called the sum of hit-density weights of the hits in the SciFi station, is used to separate signal from background. This variable was validated using hadron test-beam data. 

\begin{figure*}[t]
    \includegraphics[width=1.0\textwidth]{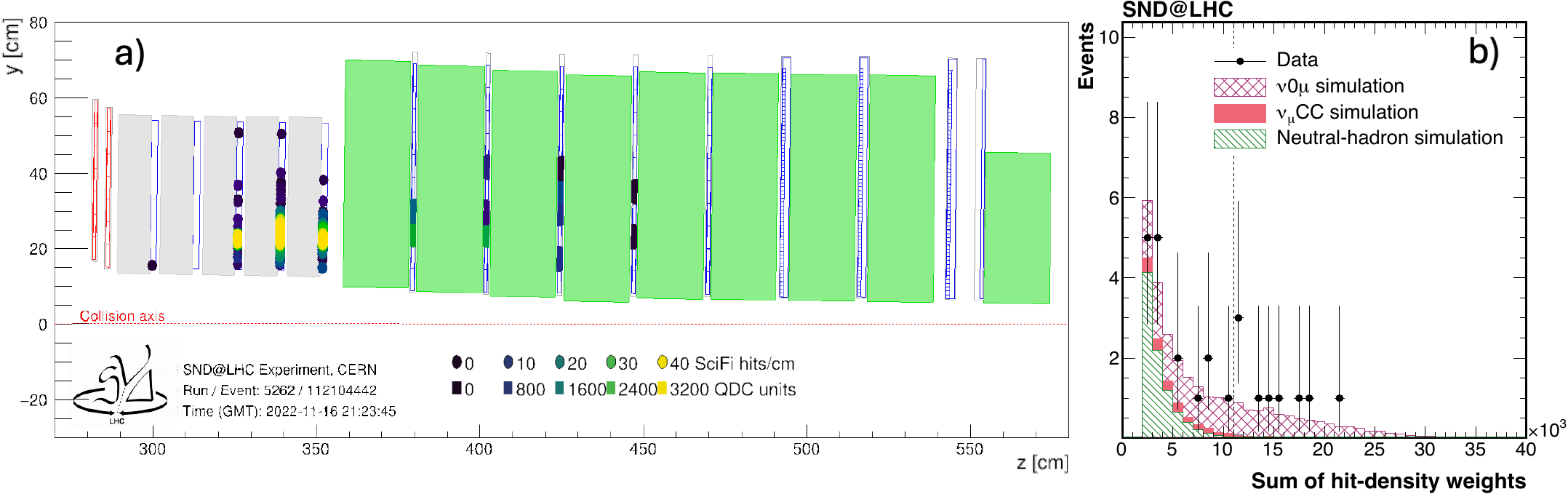}
    \caption{\textbf{a) Representative example of a signal-like 
    $\nu0\mu$ event.} The colored circles represent the local density of hits in the SciFi detector, corresponding to the number of hits within 1 cm of each hit. The colored rectangles represent the amplitude, in arbitrary units, of hits in the US hadron calorimeter. Lighter shades correspond to higher values.
    \textbf{b) Distribution of the sum of SciFi hit-density weights for events selected into the analysis sample.} The events from the data are shown alongside the expected signal and background. Figure adapted from Ref.~\cite{SNDLHC:2024qqb} (CC BY 4.0).}
    \label{fig:eventDisplay}
\end{figure*}

The expected significance of the $\nu0\mu$ observation is 5.5~$\sigma$, with 7.2 signal events expected over a background of 0.30 $\nu_\mu$ CC events, $1.5\times 10^{-2}$ neutral hadron events resulting from muon DIS interactions in the tunnel walls, and $1.7 \times 10^{-3}$ $\nu_\tau$ CC interactions with a muon in the final state. The signal expectation is composed of 4.9 $\nu_e$ CC events, 2.2 NC events, and 0.1 $\nu_\tau$ CC events with no muons in the final state.

Nine events are observed in the signal region, resulting in an observation significance of 6.4~$\sigma$. An example of a signal-like event is shown in Figure~\ref{fig:eventDisplay} (a). The distribution of the sum of SciFi hit-density weights for events observed in the data is shown in Figure~\ref{fig:eventDisplay} (b), along with signal and background expectations. Events are selected in the signal region which have a sum of SciFi hit-density weights above $11 \times 10^3$. 

The significance to observe $\nu_e$ CC interactions with this analysis is also evaluated. The NC component of the $\nu0\mu$ signal originates primarily from the flux of $\nu_\mu$, hence the NC component of the $\nu0\mu$ signal can be constrained using the measurement of $\nu_\mu$ CC interactions in Ref.~\cite{SNDLHC:2023pun}. With this constraint, evidence for the observation of $\nu_e$ CC events is extracted at the level of 3.7~$\sigma$, with an expected significance of 2.2~$\sigma$. The uncertainties on the $\nu_\mu$ CC and NC components are conservatively taken to be fully correlated. Hence the data show evidence of $\nu_e$ CC interactions in the detector. In the near future a parallel analysis based on the emulsion data on such events is expected to be
reported.

\section{FUTURE PHYSICS PROSPECTS}

\subsection{Prospects for Run 3}

Both FASER and SND@LHC will continue to take data for the remaining period of Run 3 (leading to a total projected dataset size of 350~fb$^{-1}$) and will repeat the analyses discussed before with the updated dataset size, and improved understanding of the detector performance, leading to important improvements in the precision. In addition to these analyses, the larger dataset and improved detector understanding will allow the experiments to carry out new measurements, for example observing neutral current neutrino interactions, observing tau neutrino CC interactions, studying neutrino induced charm production, and constraining lepton flavor violation effects in the neutrino sector. 

One aspect of the FASER Run 3 experimental program that has not yet been achieved is to match $\nu_\mu$ CC candidates reconstructed in the emulsion detector, to the electronic detector. This would enable the produced muon's momentum and charge to be measured with the spectrometer, and would allow separate measurements of neutrino and antineutrinos. The spectrometer acceptance in the transverse plane only covers 42\% of the FASER$\nu$ emulsion detector, so this can only be done for part of the available statistics, but will still provide important additional information. 

In SND@LHC, for the analysis of the 2022 data the imposed fiducial volume cut was found to reject 92.4\% of the neutrino CC interactions. This was mostly to mitigate the large observed veto inefficiency in the bottom part of the detector, where the neutrino density is rather high.
To remedy this deficiency, an additional third veto station with vertical bars was installed during the Year End Technical Stop of 2023–2024. The acceptance is further increased by the excavation and the shift of the whole veto system downwards. The new position of the veto now provides full coverage of the target sensitive area. Since the lower part of the target has the highest neutrino density, an increase in the number of observed neutrino interactions by about a factor of two is expected with the deployment of this extra veto layer, and correspondingly enhanced fiducial volume. The new veto detector was tested with cosmic rays and was included in the data taking from the start of the run in 2024.

\begin{figure}[hb]
\centering
\includegraphics[width=1.\textwidth]{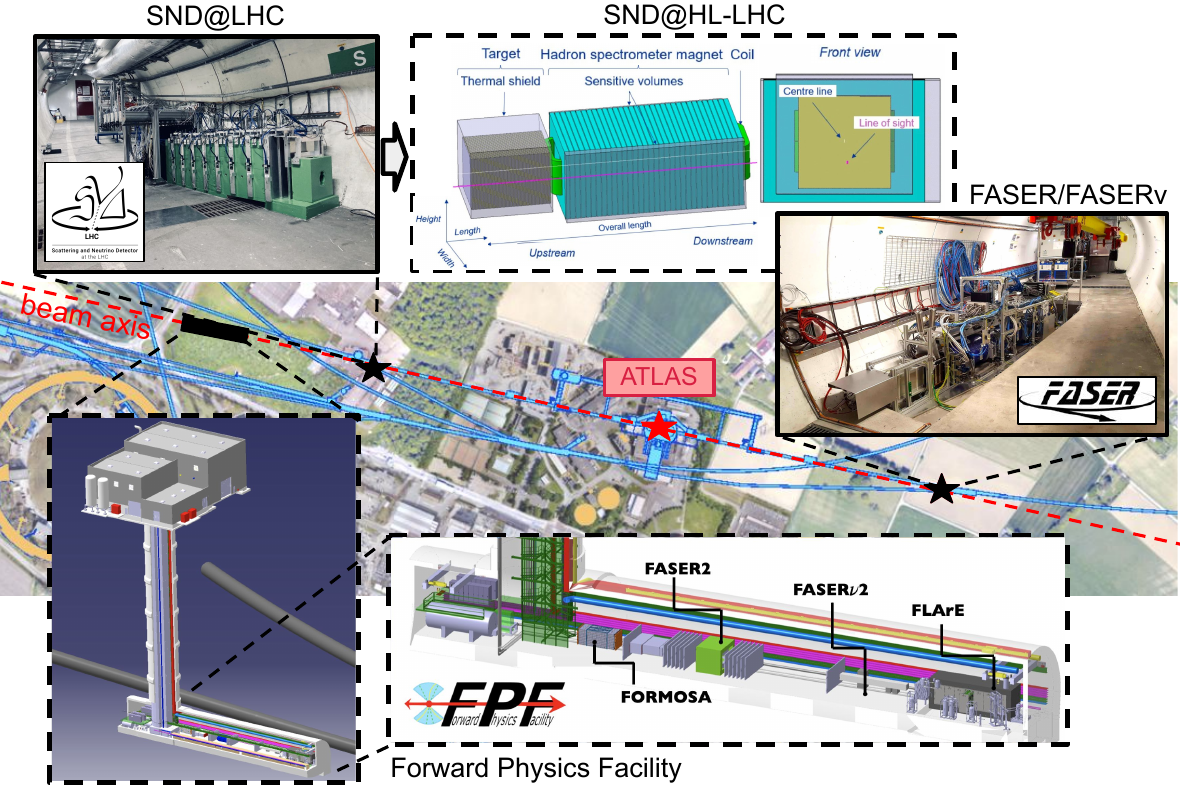}
\caption{\textbf{Existing and proposed projects.} Location of the existing and proposed LHC neutrino experiments along the beam collision axis of the ATLAS experiment (red dashed line). The inset shows  photos of the existing FASER and SND@LHC experiments, the conceptual design of the SND@LHC Phase II detector, and the layout of the FPF cavern. 
}
\label{fig:future}
\end{figure}

\subsection{FASER Run 4 upgrade}

FASER will continue taking data in LHC Run 4, but with the increased luminosity at the HL-LHC it will become difficult to efficiently use the FASER$\nu$ emulsion detector with the associated increase in the background muon rate. Along with efforts to improve FASER$\nu$ reconstruction to tolerate the high background rate, the FASER Collaboration is therefore investigating several options for installing a $\mathcal{O}$(1 tonne) target mass electronic neutrino detector that could operate in Run 4. The space currently occupied by FASER$\nu$ is in a trench that was excavated to align the detector with the beam collision axis, and would allow a new detector to be positioned in an on-axis configuration without additional civil engineering works. The main physics goal of the upgraded detector would be to be able to identify muon and electron neutrino CC interactions, to separate these from NC interactions, and to measure the neutrino energy with good resolution. A more optimistic goal would be to be able to identify and measure tau neutrino CC interactions, which is very challenging with an electronic detector, but is under study.

\subsection{SND@LHC Phase II detector}

A significant upgrade of the detector is proposed for LHC Run 4 and Run 5, in order to better exploit the potential of the future HL-LHC collider upgrade~\cite{Abbaneo:2895224, Abbaneo:2909524}. The SND@LHC Phase II detector will be in the same location as the current  SND@LHC detector, but largely extended in its capabilities. The use of nuclear emulsions is not considered to be an optimal option, given the foreseen five-fold increase in instantaneous luminosity for these runs. Furthermore, the collaboration seeks to implement structural design changes that had been impossible to implement for Run~3 due to the short time available for detector installation. In fact, the current configuration of the tunnel (sloping floor and available space) has imposed severe limitations on the SND@LHC acceptance and performance so far. The foreseen layout for the SND@LHC Phase II detector is shown schematically in Figure~\ref{fig:future}.

It is proposed to reuse the silicon microstrip sensors from the two outermost layers of the CMS Tracker Outer Barrel~\cite{CMS:2009bxg} that will be decommissioned after Run 3, to build the new target and electromagnetic calorimeter, as well as to equip the magnetized HCAL and muon system in order to achieve excellent muon momentum resolution and charge identification performance. The silicon strips provide a resolution of about 30 $\mu$m.
The total mass of the tungsten target will be about 1.3~tons, which is 60\% larger than the one used in Run~3.
The SND target station itself will consist of 58 alternating layers of $40 \times 40$~cm$^2$ tungsten absorbers and silicon trackers.
Currently detailed studies are ongoing on the efficiency of $\nu_{\tau}$ interaction identification with this detector configuration.

The aim is for the pseudo-rapidity coverage of the new detector, 
although off-axis, to include the beam spot in its acceptance, as shown in Figure~\ref{fig:future}.
The proposed HCAL design consists of 34 magnetized iron slabs, each 5~cm thick. A magnetic field strength of about 1.7~T can be achieved with a  magnetized length of 1.70~m.  
There will be enough silicon modules available to equip all active layers of the HCAL/muon system to allow the assembly of planes with a required size of $40 \times 40$ cm$^2$, enabling high-precision measurements of the muons' charge and momentum. The overall magnetised iron length of the detector will allow for an excellent performance in terms of muon momentum resolution and ensure at least 10 interaction lengths for the hadronic calorimeter. 
Fast detector planes using either plastic scintillator or resistive-plate chamber technology will be used to trigger the read out. 

The proposed detector is expected  to be able to collect $\mathcal{O}$(100k) neutrino CC events of which a few 10k events are produced in decays of charm quark flavored hadrons, and $\mathcal{O}$(40k)  NC events for a luminosity of 3 ab$^{-1}$. About 1k tau neutrino CC events will be produced in the new neutrino target. So the overall increase of statistics with respect to that expected for Run~3 is huge, amounting to a factor of 100
as shown in Fig.~\ref{fig:spectra}.

\subsection{Forward Physics Facility}

The FASER and SND@LHC detectors are located in tunnels remaining from the LEP era, which were never intended to host experiments. In particular, the available space at these locations is limited by the tunnel size and geometry, restricting the possible target mass and technologies for collider neutrino detectors. The possibility of increasing the available space with large-scale civil engineering has been investigated but was found not to be possible. 

To overcome this limitation, the Forward Physics Facility (FPF)~\cite{Adhikary:2024nlv} has been proposed and would allow multiple $\mathcal{O}(10)$ tonnes target mass detectors to be placed on the collision axis line of sight. As shown in Figure~\ref{fig:future}, the FPF is housed in a new cavern at the depth of the LHC tunnel, 617~m to the west of the ATLAS interaction point, on CERN land in France. Technical studies on the implementation of the FPF have been carried out in the context of the CERN Physics Beyond Colliders study group, and have shown that such a facility is feasible~\cite{PBCnote,PBCnote2,vibration-note}. Two neutrino detectors have been proposed for the FPF, FASER$\nu$2 which is a 20~tonne scaled up version of the tungsten-emulsion FASER$\nu$ detector, and a liquid argon Time Projection Chamber detector (FLArE) with a 10~tonne fiducial volume. The facility would also include FASER2, a large aperture tracking spectrometer, to search for long-lived particles, but also to measure the momentum and charge of muons produced in neutrino interactions in the upstream neutrino detectors. During the HL-LHC era these experiments are expected to record $\mathcal{O}(10^6)$ $\nu_\mu$, $\mathcal{O}(10^5)$ $\nu_e$, and $\mathcal{O}(10^4)$ $\nu_\tau$ interactions, maximising the physics potential of the LHC neutrino beam.  On top of the neutrino physics program the FPF has a strong BSM program, including searching for new particle scattering in FLArE, searches for the decay of new long-lived particles in FASER2, and searches for milli-charged particles in the FORMOSA experiment~\cite{Foroughi-Abari:2020qar}.

A possible timeline for the FPF would be for the civil engineering work and outfitting of the facility to happen in time for the experiments to be installed and take first data during LHC Run 4 (scheduled to finish at the end of 2033).

\section{CONCLUSIONS}

The first observation of LHC neutrinos by the FASER and SND@LHC experiments, heralds the dawn of collider neutrino physics. Following this first landmark in 2023, the experiments have continued to release updated results, with the first collider electron neutrino interactions observed, and first cross section and flux measurements. Current results have utilized data from either the electronic detector components, or a small fraction $\mathcal{O}$(1\%) of the emulsion detector data taken so far. Future results will also combine information from these systems to enhance the measurements. During the remaining period of LHC Run 3 (until mid-2026) a total dataset corresponding to about 350~fb$^{-1}$ is expected to be recorded, which will allow improved precision and several other important measurements to be made, including the first observation of tau neutrino interactions. The experiments have demonstrated that neutrino physics is possible in the harsh conditions of the LHC hadron collisions.

In recent years it has become clear that collider neutrino measurements have a broad potential for neutrino physics and to probe QCD. Neutrino scattering measurements will provide novel tests of lepton universality and constrain proton structure, while neutrino flux measurements provide an innovative method to probe forward particle production and constrain the strong interaction in a previously uncharted kinematic regime, for example allowing to probe intrinsic charm, test gluon saturation, and measure the gluon content of the proton at ultra-low momentum fractions. These measurement also provide important input for astroparticle physics, for example to validate interaction cross section models, resolve the muon puzzle in cosmic ray physics or to improve flux predictions for prompt atmospheric neutrinos. Finally, these experiments will constrain a variety of new physics scenarios in the neutrino sector and perform searches for light dark sector particles. 

Based on the successful first measurements by FASER and SND@LHC, new projects are being explored for future running at the LHC and beyond. The SND@LHC Collaboration proposes a significant upgrade in the same location as the existing experiment. This upgrade would allow a larger target mass, and to be located more optimally with respect to the beam collision axis line of sight, as well as a magnetized hadronic calorimeter and muon system. Likewise, the FASER Collaboration is exploring options for a new neutrino detector at the HL-LHC. More ambitiously, the Forward Physics Facility is a proposed new cavern which could house multiple $\mathcal{O}(10)$ tonne neutrino detectors to maximally exploit the LHC neutrino beam in the HL-LHC era. In addition, there are new, less well developed, ideas being pursued, for example detecting LHC neutrinos at further locations, such as in lake Geneva~\cite{Ariga:2025jgv,Kamp:2025phs}. In addition, collider neutrinos are now considered for future collider projects, such as at the muon collider~\cite{InternationalMuonCollider:2024jyv,Bojorquez-Lopez:2024bsr} and at future hadron colliders~\cite{MammenAbraham:2024gun}.

The collider neutrino program is still in its infancy but has already produced interesting measurements in a previously unexplored energy regime. It will be fascinating to see how this field develops over the coming years, and what new insights can be learned from studying high-energy neutrinos of all flavors produced at particle colliders.

\section*{DISCLOSURE STATEMENT}
The authors are not aware of any affiliations, memberships, funding, or financial holdings that might be perceived as affecting the objectivity of this review. 

\section*{ACKNOWLEDGMENTS}
We acknowledge useful discussions with the FASER and SND@LHC Collaborations and the Forward Physics Facility coordination group. Special thanks to Cristovao Vilela, Fabio Alicante and Giovanni De Lellis for useful discussions. A.A. is supported by the European Research Council (ERC) under the European Union’s Horizon 2020 research and innovation program (Grant agreement No. 101002690) and JSPS KAKENHI Grant No. 23H00103. F.K. is supported by the Deutsche Forschungsgemeinschaft under Germany's Excellence Strategy -- EXC 2121 Quantum Universe -- 390833306. 

\bibliography{references}

\providecommand{\href}[2]{#2}\begingroup\raggedright\begin{thebibliography}{100}

\bibitem{DeRujula:1984ns}
A.~De~Rujula, \emph{{Neutrino Physics at Future Colliders}},  in \emph{{Prague 1984, Proceedings, Trends in Physics, Vol.~1, 236-245.}}, 1984.

\bibitem{Winter:1990ry}
K.~Winter, \emph{{Detection of the tau-neutrino at the LHC}},  in \emph{{ECFA Large Hadron Collider (LHC) Workshop: Physics and Instrumentation}}, pp.~37--49, 1990, \href{https://doi.org/10.5170/CERN-1990-010-V-2.37}{DOI}.

\bibitem{DeRujula:1992sn}
A.~De~Rujula, E.~Fernandez and J.J.~Gomez-Cadenas, \emph{{Neutrino fluxes at future hadron colliders}}, \href{https://doi.org/10.1016/0550-3213(93)90427-Q}{\emph{Nucl. Phys.} {\bfseries B405} (1993) 80}.

\bibitem{Vannucci:1993ud}
F.~Vannucci, \emph{{Neutrino physics at LHC/SSC}},  Tech. Rep. \href{https://cds.cern.ch/record/253670}{LPNHE-93-03}, Paris 6. Lab. Phys. Nucl. Théor. Hautes Enérg., Paris (Aug, 1993).

\bibitem{Park:2011gh}
H.~Park, \emph{{The estimation of neutrino fluxes produced by proton-proton collisions at $\sqrt{s}=14$ TeV of the LHC}}, \href{https://doi.org/10.1007/JHEP10(2011)092}{\emph{JHEP} {\bfseries 10} (2011) 092} [\href{https://arxiv.org/abs/1110.1971}{{\ttfamily 1110.1971}}].

\bibitem{Foldenauer:2021gkm}
P.~Foldenauer, F.~Kling and P.~Reimitz, \emph{{Potential of CMS as a high-energy neutrino scattering experiment}}, \href{https://doi.org/10.1103/PhysRevD.104.113005}{\emph{Phys. Rev. D} {\bfseries 104} (2021) 113005} [\href{https://arxiv.org/abs/2108.05370}{{\ttfamily 2108.05370}}].

\bibitem{Feng:2017uoz}
J.L.~Feng, I.~Galon, F.~Kling and S.~Trojanowski, \emph{{ForwArd Search ExpeRiment at the LHC}}, \href{https://doi.org/10.1103/PhysRevD.97.035001}{\emph{Phys. Rev. D} {\bfseries 97} (2018) 035001} [\href{https://arxiv.org/abs/1708.09389}{{\ttfamily 1708.09389}}].

\bibitem{FASER:2018ceo}
{\scshape FASER} collaboration, \emph{{Letter of Intent for FASER: ForwArd Search ExpeRiment at the LHC}},  \href{https://arxiv.org/abs/1811.10243}{{\ttfamily 1811.10243}}.

\bibitem{FASER:2018bac}
{\scshape FASER} collaboration, \emph{{Technical Proposal for FASER: ForwArd Search ExpeRiment at the LHC}},  \href{https://arxiv.org/abs/1812.09139}{{\ttfamily 1812.09139}}.

\bibitem{FASER:2018eoc}
{\scshape FASER} collaboration, \emph{{FASER\textquoteright{}s physics reach for long-lived particles}}, \href{https://doi.org/10.1103/PhysRevD.99.095011}{\emph{Phys. Rev. D} {\bfseries 99} (2019) 095011} [\href{https://arxiv.org/abs/1811.12522}{{\ttfamily 1811.12522}}].

\bibitem{FASER:2019aik}
{\scshape FASER} collaboration, \emph{{FASER: ForwArd Search ExpeRiment at the LHC}},  \href{https://arxiv.org/abs/1901.04468}{{\ttfamily 1901.04468}}.

\bibitem{FASER:2021mtu}
{\scshape FASER} collaboration, \emph{{First neutrino interaction candidates at the LHC}}, \href{https://doi.org/10.1103/PhysRevD.104.L091101}{\emph{Phys. Rev. D} {\bfseries 104} (2021) L091101} [\href{https://arxiv.org/abs/2105.06197}{{\ttfamily 2105.06197}}].

\bibitem{FASER:2019dxq}
{\scshape FASER} collaboration, \emph{{Detecting and Studying High-Energy Collider Neutrinos with FASER at the LHC}}, \href{https://doi.org/10.1140/epjc/s10052-020-7631-5}{\emph{Eur. Phys. J. C} {\bfseries 80} (2020) 61} [\href{https://arxiv.org/abs/1908.02310}{{\ttfamily 1908.02310}}].

\bibitem{FASER:2020gpr}
{\scshape FASER} collaboration, \emph{{Technical Proposal: FASERnu}},  \href{https://arxiv.org/abs/2001.03073}{{\ttfamily 2001.03073}}.

\bibitem{Buontempo:2018gta}
S.~Buontempo, G.M.~Dallavalle, G.~De~Lellis, D.~Lazic and F.L.~Navarria, \emph{{CMS-XSEN: LHC Neutrinos at CMS. Experiment Feasibility Study}},  \href{https://arxiv.org/abs/1804.04413}{{\ttfamily 1804.04413}}.

\bibitem{Beni:2019gxv}
N.~Beni et~al., \emph{{Physics Potential of an Experiment using LHC Neutrinos}}, \href{https://doi.org/10.1088/1361-6471/ab3f7c}{\emph{J. Phys. G} {\bfseries 46} (2019) 115008} [\href{https://arxiv.org/abs/1903.06564}{{\ttfamily 1903.06564}}].

\bibitem{Beni:2020yfy}
N.~Beni et~al., \emph{{Further studies on the physics potential of an experiment using LHC neutrinos}}, \href{https://doi.org/10.1088/1361-6471/aba7ad}{\emph{J. Phys. G} {\bfseries 47} (2020) 125004} [\href{https://arxiv.org/abs/2004.07828}{{\ttfamily 2004.07828}}].

\bibitem{XSEN:2019bel}
{\scshape XSEN} collaboration, \emph{{XSEN: a $\nu$N Cross Section Measurement using High Energy Neutrinos from pp collisions at the LHC}},  \href{https://arxiv.org/abs/1910.11340}{{\ttfamily 1910.11340}}.

\bibitem{SHiP:2020sos}
{\scshape SHiP} collaboration, \emph{{SND@LHC}},  \href{https://arxiv.org/abs/2002.08722}{{\ttfamily 2002.08722}}.

\bibitem{Ahdida:2750060}
C.~Ahdida et~al., ``{SND@LHC - Scattering and Neutrino Detector at the LHC}.'' \url{https://cds.cern.ch/record/2750060}, Jan, 2021.

\bibitem{FASER:2023zcr}
{\scshape FASER} collaboration, \emph{{First Direct Observation of Collider Neutrinos with FASER at the LHC}}, \href{https://doi.org/10.1103/PhysRevLett.131.031801}{\emph{Phys. Rev. Lett.} {\bfseries 131} (2023) 031801} [\href{https://arxiv.org/abs/2303.14185}{{\ttfamily 2303.14185}}].

\bibitem{SNDLHC:2023pun}
{\scshape SND@LHC} collaboration, \emph{{Observation of Collider Muon Neutrinos with the SND@LHC Experiment}}, \href{https://doi.org/10.1103/PhysRevLett.131.031802}{\emph{Phys. Rev. Lett.} {\bfseries 131} (2023) 031802} [\href{https://arxiv.org/abs/2305.09383}{{\ttfamily 2305.09383}}].

\bibitem{Boyd:2882503}
{\scshape FASER} collaboration, ``{Request to run FASER in Run 4}.'' \url{https://cds.cern.ch/record/2882503}, 2023.

\bibitem{Abbaneo:2895224}
{\scshape SND@LHC} collaboration, D.~Abbaneo et~al., ``{AdvSND, The Advanced Scattering and Neutrino Detector at High Lumi LHC Letter of Intent}.'' \url{https://cds.cern.ch/record/2895224}, 2024.

\bibitem{Abbaneo:2909524}
{\scshape SND@LHC} collaboration, D.~Abbaneo et~al., ``{Addendum to the AdvancedSND LoI}.'' \url{https://cds.cern.ch/record/2909524}, 2024.

\bibitem{MammenAbraham:2020hex}
R.~Mammen~Abraham et~al., ``{Forward Physics Facility - Snowmass 2021 Letter of Interest}.'' \url{https://doi.org/10.5281/zenodo.4059893}, 2020.

\bibitem{Anchordoqui:2021ghd}
L.A.~Anchordoqui et~al., \emph{{The Forward Physics Facility: Sites, experiments, and physics potential}}, \href{https://doi.org/10.1016/j.physrep.2022.04.004}{\emph{Phys. Rept.} {\bfseries 968} (2022) 1} [\href{https://arxiv.org/abs/2109.10905}{{\ttfamily 2109.10905}}].

\bibitem{Feng:2022inv}
J.L.~Feng et~al., \emph{{The Forward Physics Facility at the High-Luminosity LHC}}, \href{https://doi.org/10.1088/1361-6471/ac865e}{\emph{J. Phys. G} {\bfseries 50} (2023) 030501} [\href{https://arxiv.org/abs/2203.05090}{{\ttfamily 2203.05090}}].

\bibitem{Adhikary:2024nlv}
J.~Adhikary et~al., \emph{{Science and Project Planning for the Forward Physics Facility in Preparation for the 2024-2026 European Particle Physics Strategy Update}},  \href{https://arxiv.org/abs/2411.04175}{{\ttfamily 2411.04175}}.

\bibitem{Ariga:2025jgv}
A.~Ariga, S.~Barwick, J.~Boyd, M.~Fieg, F.~Kling, T.~M\"akel\"a et~al., \emph{{Detecting LHC Neutrinos at Surface Level}},  \href{https://arxiv.org/abs/2501.06142}{{\ttfamily 2501.06142}}.

\bibitem{Kamp:2025phs}
N.W.~Kamp, C.A.~Arg\"uelles, A.~Karle, J.~Thomas and T.~Yuan, \emph{{Lake- and Surface-Based Detectors for Forward Neutrino Physics}},  \href{https://arxiv.org/abs/2501.08278}{{\ttfamily 2501.08278}}.

\bibitem{Worcester:2023njy}
E.~Worcester, \emph{{The Dawn of Collider Neutrino Physics}}, \href{https://doi.org/10.1103/Physics.16.113}{\emph{APS Physics} {\bfseries 16} (2023) 113}.

\bibitem{Aberle:2839677}
{\scshape SHiP} collaboration, O.~Aberle et~al., ``{BDF/SHiP at the ECN3 high-intensity beam facility}.'' \url{https://cds.cern.ch/record/2839677}, 2022.

\bibitem{Ahdida:2023okr}
C.~Ahdida et~al., \emph{{Post-LS3 Experimental Options in ECN3}},  \href{https://arxiv.org/abs/2310.17726}{{\ttfamily 2310.17726}}.

\bibitem{Candido:2023utz}
A.~Candido, A.~Garcia, G.~Magni, T.~Rabemananjara, J.~Rojo and R.~Stegeman, \emph{{Neutrino Structure Functions from GeV to EeV Energies}}, \href{https://doi.org/10.1007/JHEP05(2023)149}{\emph{JHEP} {\bfseries 05} (2023) 149} [\href{https://arxiv.org/abs/2302.08527}{{\ttfamily 2302.08527}}].

\bibitem{Jeong:2023hwe}
Y.S.~Jeong and M.H.~Reno, \emph{{Neutrino cross sections: Interface of shallow- and deep-inelastic scattering for collider neutrinos}}, \href{https://doi.org/10.1103/PhysRevD.108.113010}{\emph{Phys. Rev. D} {\bfseries 108} (2023) 113010} [\href{https://arxiv.org/abs/2307.09241}{{\ttfamily 2307.09241}}].

\bibitem{DONuT:2007bsg}
{\scshape DONuT} collaboration, \emph{{Final tau-neutrino results from the DONuT experiment}}, \href{https://doi.org/10.1103/PhysRevD.78.052002}{\emph{Phys. Rev. D} {\bfseries 78} (2008) 052002} [\href{https://arxiv.org/abs/0711.0728}{{\ttfamily 0711.0728}}].

\bibitem{OPERA:2018nar}
{\scshape OPERA} collaboration, \emph{{Final Results of the OPERA Experiment on $\nu_\tau$ Appearance in the CNGS Neutrino Beam}}, \href{https://doi.org/10.1103/PhysRevLett.120.211801}{\emph{Phys. Rev. Lett.} {\bfseries 120} (2018) 211801} [\href{https://arxiv.org/abs/1804.04912}{{\ttfamily 1804.04912}}].

\bibitem{MammenAbraham:2022xoc}
R.~Mammen~Abraham et~al., \emph{{Tau neutrinos in the next decade: from GeV to EeV}}, \href{https://doi.org/10.1088/1361-6471/ac89d2}{\emph{J. Phys. G} {\bfseries 49} (2022) 110501} [\href{https://arxiv.org/abs/2203.05591}{{\ttfamily 2203.05591}}].

\bibitem{Falkowski:2021bkq}
A.~Falkowski, M.~Gonz\'alez-Alonso, J.~Kopp, Y.~Soreq and Z.~Tabrizi, \emph{{EFT at FASER\ensuremath{\nu}}}, \href{https://doi.org/10.1007/JHEP10(2021)086}{\emph{JHEP} {\bfseries 10} (2021) 086} [\href{https://arxiv.org/abs/2105.12136}{{\ttfamily 2105.12136}}].

\bibitem{Cruz-Martinez:2023sdv}
J.M.~Cruz-Martinez, M.~Fieg, T.~Giani, P.~Krack, T.~M\"akel\"a, T.R.~Rabemananjara et~al., \emph{{The LHC as a Neutrino-Ion Collider}}, \href{https://doi.org/10.1140/epjc/s10052-024-12665-1}{\emph{Eur. Phys. J. C} {\bfseries 84} (2024) 369} [\href{https://arxiv.org/abs/2309.09581}{{\ttfamily 2309.09581}}].

\bibitem{Accardi:2012qut}
A.~Accardi et~al., \emph{{Electron Ion Collider: The Next QCD Frontier}: {Understanding the glue that binds us all}}, \href{https://doi.org/10.1140/epja/i2016-16268-9}{\emph{Eur. Phys. J. A} {\bfseries 52} (2016) 268} [\href{https://arxiv.org/abs/1212.1701}{{\ttfamily 1212.1701}}].

\bibitem{Berge:1987zw}
J.P.~Berge et~al., \emph{{Total Neutrino and Anti-neutrino Charged Current Cross-section Measurements in 100-{GeV}, 160-{GeV} and 200-{GeV} Narrow Band Beams}}, \href{https://doi.org/10.1007/BF01596895}{\emph{Z. Phys. C} {\bfseries 35} (1987) 443}.

\bibitem{NuTeV:2005wsg}
{\scshape NuTeV} collaboration, \emph{{Precise measurement of neutrino and anti-neutrino differential cross sections}}, \href{https://doi.org/10.1103/PhysRevD.74.012008}{\emph{Phys. Rev. D} {\bfseries 74} (2006) 012008} [\href{https://arxiv.org/abs/hep-ex/0509010}{{\ttfamily hep-ex/0509010}}].

\bibitem{Hammou:2024xuj}
E.~Hammou and M.~Ubiali, \emph{{Unravelling New Physics Signals at the HL-LHC with Low-Energy Constraints}},  \href{https://arxiv.org/abs/2410.00963}{{\ttfamily 2410.00963}}.

\bibitem{Wilkinson:2023vvu}
C.~Wilkinson and A.~Garcia~Soto, \emph{{Low-\ensuremath{\nu} method with LHC neutrinos}}, \href{https://doi.org/10.1103/PhysRevD.109.033010}{\emph{Phys. Rev. D} {\bfseries 109} (2024) 033010} [\href{https://arxiv.org/abs/2310.06520}{{\ttfamily 2310.06520}}].

\bibitem{Pierog:2013ria}
T.~Pierog, I.~Karpenko, J.M.~Katzy, E.~Yatsenko and K.~Werner, \emph{{EPOS LHC: Test of collective hadronization with data measured at the CERN Large Hadron Collider}}, \href{https://doi.org/10.1103/PhysRevC.92.034906}{\emph{Phys. Rev. C} {\bfseries 92} (2015) 034906} [\href{https://arxiv.org/abs/1306.0121}{{\ttfamily 1306.0121}}].

\bibitem{Ostapchenko:2010vb}
S.~Ostapchenko, \emph{{Monte Carlo treatment of hadronic interactions in enhanced Pomeron scheme: I. QGSJET-II model}}, \href{https://doi.org/10.1103/PhysRevD.83.014018}{\emph{Phys. Rev. D} {\bfseries 83} (2011) 014018} [\href{https://arxiv.org/abs/1010.1869}{{\ttfamily 1010.1869}}].

\bibitem{Riehn:2019jet}
F.~Riehn, R.~Engel, A.~Fedynitch, T.K.~Gaisser and T.~Stanev, \emph{{Hadronic interaction model Sibyll 2.3d and extensive air showers}}, \href{https://doi.org/10.1103/PhysRevD.102.063002}{\emph{Phys. Rev. D} {\bfseries 102} (2020) 063002} [\href{https://arxiv.org/abs/1912.03300}{{\ttfamily 1912.03300}}].

\bibitem{Fieg:2023kld}
M.~Fieg, F.~Kling, H.~Schulz and T.~Sj\"ostrand, \emph{{Tuning pythia for forward physics experiments}}, \href{https://doi.org/10.1103/PhysRevD.109.016010}{\emph{Phys. Rev. D} {\bfseries 109} (2024) 016010} [\href{https://arxiv.org/abs/2309.08604}{{\ttfamily 2309.08604}}].

\bibitem{LHCf:2017fnw}
{\scshape LHCf} collaboration, \emph{{Measurement of forward photon production cross-section in proton\textendash{}proton collisions at $\sqrt{s}$ = 13 TeV with the LHCf detector}}, \href{https://doi.org/10.1016/j.physletb.2017.12.050}{\emph{Phys. Lett. B} {\bfseries 780} (2018) 233} [\href{https://arxiv.org/abs/1703.07678}{{\ttfamily 1703.07678}}].

\bibitem{LHCf:2018gbv}
{\scshape LHCf} collaboration, \emph{{Measurement of inclusive forward neutron production cross section in proton-proton collisions at $ \sqrt{s}=13 $ TeV with the LHCf Arm2 detector}}, \href{https://doi.org/10.1007/JHEP11(2018)073}{\emph{JHEP} {\bfseries 11} (2018) 073} [\href{https://arxiv.org/abs/1808.09877}{{\ttfamily 1808.09877}}].

\bibitem{FASER:2021pkt}
{\scshape FASER} collaboration, \emph{{Studying neutrinos at the LHC: FASER and its impact to the cosmic-ray physics}}, \href{https://doi.org/10.22323/1.395.1025}{\emph{PoS} {\bfseries ICRC2021} (2021) 1025}.

\bibitem{PierreAuger:2014ucz}
{\scshape Pierre Auger} collaboration, \emph{{Muons in Air Showers at the Pierre Auger Observatory: Mean Number in Highly Inclined Events}}, \href{https://doi.org/10.1103/PhysRevD.91.032003}{\emph{Phys. Rev. D} {\bfseries 91} (2015) 032003} [\href{https://arxiv.org/abs/1408.1421}{{\ttfamily 1408.1421}}].

\bibitem{PierreAuger:2016nfk}
{\scshape Pierre Auger} collaboration, \emph{{Testing Hadronic Interactions at Ultrahigh Energies with Air Showers Measured by the Pierre Auger Observatory}}, \href{https://doi.org/10.1103/PhysRevLett.117.192001}{\emph{Phys. Rev. Lett.} {\bfseries 117} (2016) 192001} [\href{https://arxiv.org/abs/1610.08509}{{\ttfamily 1610.08509}}].

\bibitem{EAS-MSU:2019kmv}
{\scshape EAS-MSU, IceCube, KASCADE-Grande, NEVOD-DECOR, Pierre Auger, SUGAR, Telescope Array, Yakutsk EAS Array} collaboration, \emph{{Report on Tests and Measurements of Hadronic Interaction Properties with Air Showers}}, \href{https://doi.org/10.1051/epjconf/201921002004}{\emph{EPJ Web Conf.} {\bfseries 210} (2019) 02004} [\href{https://arxiv.org/abs/1902.08124}{{\ttfamily 1902.08124}}].

\bibitem{Soldin:2021wyv}
{\scshape EAS-MSU, IceCube, KASCADE-Grande, NEVOD-DECOR, Pierre Auger, SUGAR, Telescope Array, Yakutsk EAS Array} collaboration, \emph{{Update on the Combined Analysis of Muon Measurements from Nine Air Shower Experiments}}, \href{https://doi.org/10.22323/1.395.0349}{\emph{PoS} {\bfseries ICRC2021} (2021) 349} [\href{https://arxiv.org/abs/2108.08341}{{\ttfamily 2108.08341}}].

\bibitem{PierreAuger:2024neu}
{\scshape Pierre Auger} collaboration, \emph{{Testing hadronic-model predictions of depth of maximum of air-shower profiles and ground-particle signals using hybrid data of the Pierre Auger Observatory}}, \href{https://doi.org/10.1103/PhysRevD.109.102001}{\emph{Phys. Rev. D} {\bfseries 109} (2024) 102001} [\href{https://arxiv.org/abs/2401.10740}{{\ttfamily 2401.10740}}].

\bibitem{Ulrich:2010rg}
R.~Ulrich, R.~Engel and M.~Unger, \emph{{Hadronic Multiparticle Production at Ultra-High Energies and Extensive Air Showers}}, \href{https://doi.org/10.1103/PhysRevD.83.054026}{\emph{Phys. Rev. D} {\bfseries 83} (2011) 054026} [\href{https://arxiv.org/abs/1010.4310}{{\ttfamily 1010.4310}}].

\bibitem{Albrecht:2021cxw}
J.~Albrecht et~al., \emph{{The Muon Puzzle in cosmic-ray induced air showers and its connection to the Large Hadron Collider}}, \href{https://doi.org/10.1007/s10509-022-04054-5}{\emph{Astrophys. Space Sci.} {\bfseries 367} (2022) 27} [\href{https://arxiv.org/abs/2105.06148}{{\ttfamily 2105.06148}}].

\bibitem{Allen:2013hfa}
J.~Allen and G.~Farrar, \emph{{Testing models of new physics with UHE air shower observations}},  in \emph{{33rd International Cosmic Ray Conference}}, p.~1182, 7, 2013 [\href{https://arxiv.org/abs/1307.7131}{{\ttfamily 1307.7131}}].

\bibitem{Anchordoqui:2016oxy}
L.A.~Anchordoqui, H.~Goldberg and T.J.~Weiler, \emph{{Strange fireball as an explanation of the muon excess in Auger data}}, \href{https://doi.org/10.1103/PhysRevD.95.063005}{\emph{Phys. Rev. D} {\bfseries 95} (2017) 063005} [\href{https://arxiv.org/abs/1612.07328}{{\ttfamily 1612.07328}}].

\bibitem{Anchordoqui:2019laz}
L.A.~Anchordoqui, C.~Garc\'\i{}a~Canal, S.J.~Sciutto and J.F.~Soriano, \emph{{Through the looking-glass with ALICE into the quark-gluon plasma: A new test for hadronic interaction models used in air shower simulations}}, \href{https://doi.org/10.1016/j.physletb.2020.135837}{\emph{Phys. Lett. B} {\bfseries 810} (2020) 135837} [\href{https://arxiv.org/abs/1907.09816}{{\ttfamily 1907.09816}}].

\bibitem{Anchordoqui:2022fpn}
L.A.~Anchordoqui, C.G.~Canal, F.~Kling, S.J.~Sciutto and J.F.~Soriano, \emph{{An explanation of the muon puzzle of ultrahigh-energy cosmic rays and the role of the Forward Physics Facility for model improvement}}, \href{https://doi.org/10.1016/j.jheap.2022.03.004}{\emph{JHEAp} {\bfseries 34} (2022) 19} [\href{https://arxiv.org/abs/2202.03095}{{\ttfamily 2202.03095}}].

\bibitem{Sciutto:2023zuz}
S.J.~Sciutto, L.A.~Anchordoqui, C.~Garcia~Canal, F.~Kling and J.F.~Soriano, \emph{{Gauging the cosmic ray muon puzzle with the Forward Physics Facility}}, \href{https://doi.org/10.22323/1.444.0388}{\emph{PoS} {\bfseries ICRC2023} (2023) 388} [\href{https://arxiv.org/abs/2307.08634}{{\ttfamily 2307.08634}}].

\bibitem{Bai:2020ukz}
W.~Bai, M.~Diwan, M.V.~Garzelli, Y.S.~Jeong and M.H.~Reno, \emph{{Far-forward neutrinos at the Large Hadron Collider}}, \href{https://doi.org/10.1007/JHEP06(2020)032}{\emph{JHEP} {\bfseries 06} (2020) 032} [\href{https://arxiv.org/abs/2002.03012}{{\ttfamily 2002.03012}}].

\bibitem{Maciula:2022lzk}
R.~Maciula and A.~Szczurek, \emph{{Far-forward production of charm mesons and neutrinos at forward physics facilities at the LHC and the intrinsic charm in the proton}}, \href{https://doi.org/10.1103/PhysRevD.107.034002}{\emph{Phys. Rev. D} {\bfseries 107} (2023) 034002} [\href{https://arxiv.org/abs/2210.08890}{{\ttfamily 2210.08890}}].

\bibitem{Bhattacharya:2023zei}
A.~Bhattacharya, F.~Kling, I.~Sarcevic and A.M.~Stasto, \emph{{Forward neutrinos from charm at the Large Hadron Collider}}, \href{https://doi.org/10.1103/PhysRevD.109.014040}{\emph{Phys. Rev. D} {\bfseries 109} (2024) 014040} [\href{https://arxiv.org/abs/2306.01578}{{\ttfamily 2306.01578}}].

\bibitem{Buonocore:2023kna}
L.~Buonocore, F.~Kling, L.~Rottoli and J.~Sominka, \emph{{Predictions for neutrinos and new physics from forward heavy hadron production at the LHC}}, \href{https://doi.org/10.1140/epjc/s10052-024-12726-5}{\emph{Eur. Phys. J. C} {\bfseries 84} (2024) 363} [\href{https://arxiv.org/abs/2309.12793}{{\ttfamily 2309.12793}}].

\bibitem{Bai:2021ira}
W.~Bai, M.~Diwan, M.V.~Garzelli, Y.S.~Jeong, F.K.~Kumar and M.H.~Reno, \emph{{Parton distribution function uncertainties in theoretical predictions for far-forward tau neutrinos at the Large Hadron Collider}}, \href{https://doi.org/10.1007/JHEP06(2022)148}{\emph{JHEP} {\bfseries 06} (2022) 148} [\href{https://arxiv.org/abs/2112.11605}{{\ttfamily 2112.11605}}].

\bibitem{Abbasi:2021qfz}
R.~Abbasi et~al., \emph{{Improved Characterization of the Astrophysical Muon\textendash{}neutrino Flux with 9.5 Years of IceCube Data}}, \href{https://doi.org/10.3847/1538-4357/ac4d29}{\emph{Astrophys. J.} {\bfseries 928} (2022) 50} [\href{https://arxiv.org/abs/2111.10299}{{\ttfamily 2111.10299}}].

\bibitem{Bai:2022xad}
W.~Bai, M.~Diwan, M.V.~Garzelli, Y.S.~Jeong, K.~Kumar and M.H.~Reno, \emph{{Forward production of prompt neutrinos from charm in the atmosphere and at high energy colliders}}, \href{https://doi.org/10.1007/JHEP10(2023)142}{\emph{JHEP} {\bfseries 10} (2023) 142} [\href{https://arxiv.org/abs/2212.07865}{{\ttfamily 2212.07865}}].

\bibitem{Anchordoqui:2023qxv}
L.A.~Anchordoqui, I.~Antoniadis, K.~Benakli, J.~Cunat and D.~Lust, \emph{{Searching for neutrino-modulino oscillations at the Forward Physics Facility}}, \href{https://doi.org/10.1016/j.physletb.2024.138530}{\emph{Phys. Lett. B} {\bfseries 850} (2024) 138530} [\href{https://arxiv.org/abs/2308.11476}{{\ttfamily 2308.11476}}].

\bibitem{Anchordoqui:2024ynb}
L.A.~Anchordoqui, I.~Antoniadis, K.~Benakli, J.~Cunat and D.~Lust, \emph{{SUSY at the FPF}},  \href{https://arxiv.org/abs/2410.16342}{{\ttfamily 2410.16342}}.

\bibitem{Kling:2020iar}
F.~Kling, \emph{{Probing light gauge bosons in tau neutrino experiments}}, \href{https://doi.org/10.1103/PhysRevD.102.015007}{\emph{Phys. Rev. D} {\bfseries 102} (2020) 015007} [\href{https://arxiv.org/abs/2005.03594}{{\ttfamily 2005.03594}}].

\bibitem{Batell:2021snh}
B.~Batell, J.L.~Feng, M.~Fieg, A.~Ismail, F.~Kling, R.M.~Abraham et~al., \emph{{Hadrophilic dark sectors at the Forward Physics Facility}}, \href{https://doi.org/10.1103/PhysRevD.105.075001}{\emph{Phys. Rev. D} {\bfseries 105} (2022) 075001} [\href{https://arxiv.org/abs/2111.10343}{{\ttfamily 2111.10343}}].

\bibitem{MammenAbraham:2023psg}
R.~Mammen~Abraham, S.~Foroughi-Abari, F.~Kling and Y.-D.~Tsai, \emph{{Neutrino Electromagnetic Properties and the Weak Mixing Angle at the LHC Forward Physics Facility}},  \href{https://arxiv.org/abs/2301.10254}{{\ttfamily 2301.10254}}.

\bibitem{Ismail:2021dyp}
A.~Ismail, S.~Jana and R.M.~Abraham, \emph{{Neutrino up-scattering via the dipole portal at forward LHC detectors}}, \href{https://doi.org/10.1103/PhysRevD.105.055008}{\emph{Phys. Rev. D} {\bfseries 105} (2022) 055008} [\href{https://arxiv.org/abs/2109.05032}{{\ttfamily 2109.05032}}].

\bibitem{Ismail:2020yqc}
A.~Ismail, R.~Mammen~Abraham and F.~Kling, \emph{{Neutral current neutrino interactions at FASER$\nu$}}, \href{https://doi.org/10.1103/PhysRevD.103.056014}{\emph{Phys. Rev. D} {\bfseries 103} (2021) 056014} [\href{https://arxiv.org/abs/2012.10500}{{\ttfamily 2012.10500}}].

\bibitem{Kling:2023tgr}
F.~Kling, T.~M\"akel\"a and S.~Trojanowski, \emph{{Investigating the fluxes and physics potential of LHC neutrino experiments}}, \href{https://doi.org/10.1103/PhysRevD.108.095020}{\emph{Phys. Rev. D} {\bfseries 108} (2023) 095020} [\href{https://arxiv.org/abs/2309.10417}{{\ttfamily 2309.10417}}].

\bibitem{Kelly:2021mcd}
K.J.~Kelly, F.~Kling, D.~Tuckler and Y.~Zhang, \emph{{Probing neutrino-portal dark matter at the Forward Physics Facility}}, \href{https://doi.org/10.1103/PhysRevD.105.075026}{\emph{Phys. Rev. D} {\bfseries 105} (2022) 075026} [\href{https://arxiv.org/abs/2111.05868}{{\ttfamily 2111.05868}}].

\bibitem{Berryman:2022hds}
J.M.~Berryman et~al., \emph{{Neutrino self-interactions: A white paper}}, \href{https://doi.org/10.1016/j.dark.2023.101267}{\emph{Phys. Dark Univ.} {\bfseries 42} (2023) 101267} [\href{https://arxiv.org/abs/2203.01955}{{\ttfamily 2203.01955}}].

\bibitem{Francener:2024wul}
R.~Francener, V.P.~Goncalves and D.R.~Gratieri, \emph{{Neutrino trident scattering at the LHC energy regime}}, \href{https://doi.org/10.1140/epjc/s10052-024-13323-2}{\emph{Eur. Phys. J. C} {\bfseries 84} (2024) 923} [\href{https://arxiv.org/abs/2406.13593}{{\ttfamily 2406.13593}}].

\bibitem{Altmannshofer:2024hqd}
W.~Altmannshofer, T.~M\"akel\"a, S.~Sarkar, S.~Trojanowski, K.~Xie and B.~Zhou, \emph{{Discovering neutrino tridents at the Large Hadron Collider}}, \href{https://doi.org/10.1103/PhysRevD.110.072018}{\emph{Phys. Rev. D} {\bfseries 110} (2024) 072018} [\href{https://arxiv.org/abs/2406.16803}{{\ttfamily 2406.16803}}].

\bibitem{Boyarsky:2021moj}
A.~Boyarsky, O.~Mikulenko, M.~Ovchynnikov and L.~Shchutska, \emph{{Searches for new physics at SND@LHC}}, \href{https://doi.org/10.1007/JHEP03(2022)006}{\emph{JHEP} {\bfseries 03} (2022) 006} [\href{https://arxiv.org/abs/2104.09688}{{\ttfamily 2104.09688}}].

\bibitem{Batell:2021blf}
B.~Batell, J.L.~Feng and S.~Trojanowski, \emph{{Detecting Dark Matter with Far-Forward Emulsion and Liquid Argon Detectors at the LHC}}, \href{https://doi.org/10.1103/PhysRevD.103.075023}{\emph{Phys. Rev. D} {\bfseries 103} (2021) 075023} [\href{https://arxiv.org/abs/2101.10338}{{\ttfamily 2101.10338}}].

\bibitem{Batell:2021aja}
B.~Batell, J.L.~Feng, A.~Ismail, F.~Kling, R.M.~Abraham and S.~Trojanowski, \emph{{Discovering dark matter at the LHC through its nuclear scattering in far-forward emulsion and liquid argon detectors}}, \href{https://doi.org/10.1103/PhysRevD.104.035036}{\emph{Phys. Rev. D} {\bfseries 104} (2021) 035036} [\href{https://arxiv.org/abs/2107.00666}{{\ttfamily 2107.00666}}].

\bibitem{Kling:2022ykt}
F.~Kling, J.-L.~Kuo, S.~Trojanowski and Y.-D.~Tsai, \emph{{FLArE up dark sectors with EM form factors at the LHC forward physics facility}}, \href{https://doi.org/10.1016/j.nuclphysb.2023.116103}{\emph{Nucl. Phys. B} {\bfseries 987} (2023) 116103} [\href{https://arxiv.org/abs/2205.09137}{{\ttfamily 2205.09137}}].

\bibitem{Kling:2022ehv}
F.~Kling and P.~Qu\'\i{}lez, \emph{{ALP searches at the LHC: FASER as a light-shining-through-walls experiment}}, \href{https://doi.org/10.1103/PhysRevD.106.055036}{\emph{Phys. Rev. D} {\bfseries 106} (2022) 055036} [\href{https://arxiv.org/abs/2204.03599}{{\ttfamily 2204.03599}}].

\bibitem{Ariga:2023fjg}
A.~Ariga, R.~Balkin, I.~Galon, E.~Kajomovitz and Y.~Soreq, \emph{{Hunting muonic forces at emulsion detectors}}, \href{https://doi.org/10.1103/PhysRevD.109.035003}{\emph{Phys. Rev. D} {\bfseries 109} (2024) 035003} [\href{https://arxiv.org/abs/2305.03102}{{\ttfamily 2305.03102}}].

\bibitem{Batell:2024cdl}
B.~Batell, H.~Davoudiasl, R.~Marcarelli, E.T.~Neil and S.~Trojanowski, \emph{{Lepton-flavor-violating ALP signals with TeV-scale muon beams}}, \href{https://doi.org/10.1103/PhysRevD.110.075039}{\emph{Phys. Rev. D} {\bfseries 110} (2024) 075039} [\href{https://arxiv.org/abs/2407.15942}{{\ttfamily 2407.15942}}].

\bibitem{FASER:2023tle}
{\scshape FASER} collaboration, \emph{{Search for dark photons with the FASER detector at the LHC}}, \href{https://doi.org/10.1016/j.physletb.2023.138378}{\emph{Phys. Lett. B} {\bfseries 848} (2024) 138378} [\href{https://arxiv.org/abs/2308.05587}{{\ttfamily 2308.05587}}].

\bibitem{FASER:2024bbl}
{\scshape FASER} collaboration, \emph{{Shining Light on the Dark Sector: Search for Axion-like Particles and Other New Physics in Photonic Final States with FASER}},  \href{https://arxiv.org/abs/2410.10363}{{\ttfamily 2410.10363}}.

\bibitem{SNDLHC:2023mib}
{\scshape SND@LHC} collaboration, \emph{{Measurement of the muon flux at the SND@LHC experiment}}, \href{https://doi.org/10.1140/epjc/s10052-023-12380-3}{\emph{Eur. Phys. J. C} {\bfseries 84} (2024) 90} [\href{https://arxiv.org/abs/2310.05536}{{\ttfamily 2310.05536}}].

\bibitem{Battistoni:2015epi}
G.~Battistoni et~al., \emph{{Overview of the FLUKA code}}, \href{https://doi.org/10.1016/j.anucene.2014.11.007}{\emph{Annals Nucl. Energy} {\bfseries 82} (2015) 10}.

\bibitem{Ariga:2020lbq}
A.~Ariga, T.~Ariga, G.~De~Lellis, A.~Ereditato and K.~Niwa, \emph{{Nuclear Emulsions}}, \href{https://doi.org/10.1007/978-3-030-35318-6_9}{\emph{{Particle Physics Reference Library. Volume 2: Detectors for Particles and Radiation}} (2020) 383}.

\bibitem{FASER:2022hcn}
{\scshape FASER} collaboration, \emph{{The FASER detector}}, \href{https://doi.org/10.1088/1748-0221/19/05/P05066}{\emph{JINST} {\bfseries 19} (2024) P05066} [\href{https://arxiv.org/abs/2207.11427}{{\ttfamily 2207.11427}}].

\bibitem{Arakawa:2022rmp}
J.~Arakawa, J.L.~Feng, A.~Ismail, F.~Kling and M.~Waterbury, \emph{{Neutrino detection without neutrino detectors: Discovering collider neutrinos at FASER with electronic signals only}}, \href{https://doi.org/10.1103/PhysRevD.106.052011}{\emph{Phys. Rev. D} {\bfseries 106} (2022) 052011} [\href{https://arxiv.org/abs/2206.09932}{{\ttfamily 2206.09932}}].

\bibitem{Yoshimoto:2017ufm}
M.~Yoshimoto, T.~Nakano, R.~Komatani and H.~Kawahara, \emph{{Hyper-track selector nuclear emulsion readout system aimed at scanning an area of one thousand square meters}}, \href{https://doi.org/10.1093/ptep/ptx131}{\emph{PTEP} {\bfseries 2017} (2017) 103H01} [\href{https://arxiv.org/abs/1704.06814}{{\ttfamily 1704.06814}}].

\bibitem{FASER:2024hoe}
{\scshape FASER} collaboration, \emph{{First Measurement of \ensuremath{\nu_e} and \ensuremath{\nu_\mu} Interaction Cross Sections at the LHC with FASER\textquoteright{}s Emulsion Detector}}, \href{https://doi.org/10.1103/PhysRevLett.133.021802}{\emph{Phys. Rev. Lett.} {\bfseries 133} (2024) 021802} [\href{https://arxiv.org/abs/2403.12520}{{\ttfamily 2403.12520}}].

\bibitem{FASER:Tracker}
{\scshape FASER} collaboration, \emph{{The tracking detector of the FASER experiment}}, \href{https://doi.org/10.1016/j.nima.2022.166825}{\emph{Nucl. Instrum. Meth. A} {\bfseries 1034} (2022) 166825} [\href{https://arxiv.org/abs/2112.01116}{{\ttfamily 2112.01116}}].

\bibitem{FASERTDAQ:2021}
{\scshape FASER} collaboration, \emph{{The Trigger and Data Acquisition System of the FASER Experiment}}, \href{https://doi.org/10.1088/1748-0221/16/12/P12028}{\emph{JINST} {\bfseries 16} (2021) P12028} [\href{https://arxiv.org/abs/2110.15186}{{\ttfamily 2110.15186}}].

\bibitem{SHiP:2015vad}
{\scshape SHiP} collaboration, \emph{{A facility to Search for Hidden Particles (SHiP) at the CERN SPS}},  \href{https://arxiv.org/abs/1504.04956}{{\ttfamily 1504.04956}}.

\bibitem{SNDLHC:2022ihg}
{\scshape SND@LHC} collaboration, \emph{{SND@LHC: the scattering and neutrino detector at the LHC}}, \href{https://doi.org/10.1088/1748-0221/19/05/P05067}{\emph{JINST} {\bfseries 19} (2024) P05067} [\href{https://arxiv.org/abs/2210.02784}{{\ttfamily 2210.02784}}].

\bibitem{Kling:2021gos}
F.~Kling and L.J.~Nevay, \emph{{Forward neutrino fluxes at the LHC}}, \href{https://doi.org/10.1103/PhysRevD.104.113008}{\emph{Phys. Rev. D} {\bfseries 104} (2021) 113008} [\href{https://arxiv.org/abs/2105.08270}{{\ttfamily 2105.08270}}].

\bibitem{FASER:2024ykc}
{\scshape FASER} collaboration, \emph{{Neutrino rate predictions for FASER}}, \href{https://doi.org/10.1103/PhysRevD.110.012009}{\emph{Phys. Rev. D} {\bfseries 110} (2024) 012009} [\href{https://arxiv.org/abs/2402.13318}{{\ttfamily 2402.13318}}].

\bibitem{Nason:2004rx}
P.~Nason, \emph{{A New method for combining NLO QCD with shower Monte Carlo algorithms}}, \href{https://doi.org/10.1088/1126-6708/2004/11/040}{\emph{JHEP} {\bfseries 11} (2004) 040} [\href{https://arxiv.org/abs/hep-ph/0409146}{{\ttfamily hep-ph/0409146}}].

\bibitem{Frixione:2007vw}
S.~Frixione, P.~Nason and C.~Oleari, \emph{{Matching NLO QCD computations with Parton Shower simulations: the POWHEG method}}, \href{https://doi.org/10.1088/1126-6708/2007/11/070}{\emph{JHEP} {\bfseries 11} (2007) 070} [\href{https://arxiv.org/abs/0709.2092}{{\ttfamily 0709.2092}}].

\bibitem{Alioli:2010xd}
S.~Alioli, P.~Nason, C.~Oleari and E.~Re, \emph{{A general framework for implementing NLO calculations in shower Monte Carlo programs: the POWHEG BOX}}, \href{https://doi.org/10.1007/JHEP06(2010)043}{\emph{JHEP} {\bfseries 06} (2010) 043} [\href{https://arxiv.org/abs/1002.2581}{{\ttfamily 1002.2581}}].

\bibitem{Bierlich:2022pfr}
C.~Bierlich et~al., \emph{{A comprehensive guide to the physics and usage of PYTHIA 8.3}}, \href{https://doi.org/10.21468/SciPostPhysCodeb.8}{\emph{SciPost Phys. Codeb.} {\bfseries 2022} (2022) 8} [\href{https://arxiv.org/abs/2203.11601}{{\ttfamily 2203.11601}}].

\bibitem{Andreopoulos:2009rq}
C.~Andreopoulos et~al., \emph{{The GENIE Neutrino Monte Carlo Generator}}, \href{https://doi.org/10.1016/j.nima.2009.12.009}{\emph{Nucl. Instrum. Meth. A} {\bfseries 614} (2010) 87} [\href{https://arxiv.org/abs/0905.2517}{{\ttfamily 0905.2517}}].

\bibitem{Bodek:2002vp}
A.~Bodek and U.K.~Yang, \emph{{Modeling deep inelastic cross-sections in the few GeV region}}, \href{https://doi.org/10.1016/S0920-5632(02)01755-3}{\emph{Nucl. Phys. B Proc. Suppl.} {\bfseries 112} (2002) 70} [\href{https://arxiv.org/abs/hep-ex/0203009}{{\ttfamily hep-ex/0203009}}].

\bibitem{Bodek:2004pc}
A.~Bodek, I.~Park and U.-k.~Yang, \emph{{Improved low Q**2 model for neutrino and electron nucleon cross sections in few GeV region}}, \href{https://doi.org/10.1016/j.nuclphysbps.2004.11.208}{\emph{Nucl. Phys. B Proc. Suppl.} {\bfseries 139} (2005) 113} [\href{https://arxiv.org/abs/hep-ph/0411202}{{\ttfamily hep-ph/0411202}}].

\bibitem{Bodek:2010km}
A.~Bodek and U.-k.~Yang, \emph{{Axial and Vector Structure Functions for Electron- and Neutrino- Nucleon Scattering Cross Sections at all $Q^2$ using Effective Leading order Parton Distribution Functions}},  \href{https://arxiv.org/abs/1011.6592}{{\ttfamily 1011.6592}}.

\bibitem{SNDLHC:2024bzp}
{\scshape SND@LHC} collaboration, \emph{{Results and Perspectives from the First Two Years of Neutrino Physics at the LHC by the SND@LHC Experiment}}, \href{https://doi.org/10.3390/sym16060702}{\emph{Symmetry} {\bfseries 16} (2024) 702}.

\bibitem{fujimori_2024_13222911}
H.~Fujimori, \emph{{Momentum measurement in the FASER$\nu$ detector in the LHC-FASER experiment}},  Aug, 2024.
\newblock 10.5281/zenodo.13222911.

\bibitem{FASER:2024ref}
{\scshape FASER} collaboration, \emph{{First Measurement of the Muon Neutrino Interaction Cross Section and Flux as a Function of Energy at the LHC with FASER}},  \href{https://arxiv.org/abs/2412.03186}{{\ttfamily 2412.03186}}.

\bibitem{NOMAD:2007krq}
{\scshape NOMAD} collaboration, \emph{{A Precise measurement of the muon neutrino-nucleon inclusive charged current cross-section off an isoscalar target in the energy range 2.5 \ensuremath{<} E(nu) \ensuremath{<} 40-GeV by NOMAD}}, \href{https://doi.org/10.1016/j.physletb.2007.12.027}{\emph{Phys. Lett. B} {\bfseries 660} (2008) 19} [\href{https://arxiv.org/abs/0711.1183}{{\ttfamily 0711.1183}}].

\bibitem{Seligman:1997fe}
W.G.~Seligman, \emph{{A Next-to-Leading Order QCD Analysis of Neutrino - Iron Structure Functions at the Tevatron}}, Ph.D. thesis, Nevis Labs, Columbia U., 1997.
\newblock 10.2172/1421736.

\bibitem{Baltay:1988au}
C.~Baltay et~al., \emph{{$\nu_\mu - \nu_e$ Universality in Charged Current Neutrino Interactions}}, \href{https://doi.org/10.1103/PhysRevD.41.2653}{\emph{Phys. Rev. D} {\bfseries 41} (1990) 2653}.

\bibitem{IceCube:2017roe}
{\scshape IceCube} collaboration, \emph{{Measurement of the multi-TeV neutrino cross section with IceCube using Earth absorption}}, \href{https://doi.org/10.1038/nature24459}{\emph{Nature} {\bfseries 551} (2017) 596} [\href{https://arxiv.org/abs/1711.08119}{{\ttfamily 1711.08119}}].

\bibitem{Bustamante:2017xuy}
M.~Bustamante and A.~Connolly, \emph{{Extracting the Energy-Dependent Neutrino-Nucleon Cross Section above 10 TeV Using IceCube Showers}}, \href{https://doi.org/10.1103/PhysRevLett.122.041101}{\emph{Phys. Rev. Lett.} {\bfseries 122} (2019) 041101} [\href{https://arxiv.org/abs/1711.11043}{{\ttfamily 1711.11043}}].

\bibitem{IceCube:2020rnc}
{\scshape IceCube} collaboration, \emph{{Measurement of the high-energy all-flavor neutrino-nucleon cross section with IceCube}},  \href{https://arxiv.org/abs/2011.03560}{{\ttfamily 2011.03560}}.

\bibitem{SNDMoriond24}
{\scshape SND@LHC} collaboration, C.~Vilela, ``{Recent results from the SND@LHC experiment [Conference presentation]}.'' \url{https://moriond.in2p3.fr/2024/QCD/}, 2024.

\bibitem{SNDLHC:2024qqb}
{\scshape SND@LHC} collaboration, \emph{{Observation of collider neutrinos without final state muons with the SND@LHC experiment}},  \href{https://arxiv.org/abs/2411.18787}{{\ttfamily 2411.18787}}.

\bibitem{CMS:2009bxg}
{\scshape CMS} collaboration, \emph{{Commissioning and Performance of the CMS Pixel Tracker with Cosmic Ray Muons}}, \href{https://doi.org/10.1088/1748-0221/5/03/T03007}{\emph{JINST} {\bfseries 5} (2010) T03007} [\href{https://arxiv.org/abs/0911.5434}{{\ttfamily 0911.5434}}].

\bibitem{PBCnote}
J.~Boyd et~al., ``{Update on the FPF Facility technical studies}.'' \url{https://cds.cern.ch/record/2851822}, 2023.

\bibitem{PBCnote2}
J.~Boyd et~al., ``{Update of Facility Technical Studies for the FPF}.'' \url{https://cds.cern.ch/record/2904086}, 2024.

\bibitem{vibration-note}
D.~Gamba et~al., ``{Impact of Vibration to HL-LHC Performance During the FPF Facility Construction}.'' \url{https://cds.cern.ch/record/2901520}, 2024.

\bibitem{Foroughi-Abari:2020qar}
S.~Foroughi-Abari, F.~Kling and Y.-D.~Tsai, \emph{{Looking forward to millicharged dark sectors at the LHC}}, \href{https://doi.org/10.1103/PhysRevD.104.035014}{\emph{Phys. Rev. D} {\bfseries 104} (2021) 035014} [\href{https://arxiv.org/abs/2010.07941}{{\ttfamily 2010.07941}}].

\bibitem{InternationalMuonCollider:2024jyv}
{\scshape International Muon Collider} collaboration, \emph{{Interim report for the International Muon Collider Collaboration (IMCC)}},  \href{https://arxiv.org/abs/2407.12450}{{\ttfamily 2407.12450}}.

\bibitem{Bojorquez-Lopez:2024bsr}
L.~Bojorquez-Lopez, M.~Hostert, C.A.~Arg\"uelles and Z.~Liu, \emph{{The Neutrino Slice at Muon Colliders}},  \href{https://arxiv.org/abs/2412.14115}{{\ttfamily 2412.14115}}.

\bibitem{MammenAbraham:2024gun}
R.~Mammen~Abraham, J.~Adhikary, J.L.~Feng, M.~Fieg, F.~Kling, J.~Li et~al., \emph{{FPF@FCC: Neutrino, QCD, and BSM Physics Opportunities with Far-Forward Experiments at a 100 TeV Proton Collider}},  \href{https://arxiv.org/abs/2409.02163}{{\ttfamily 2409.02163}}.

\end{thebibliography}\endgroup

\end{document}